\documentclass[gmd, manuscript]{copernicus}

\usepackage[utf8]{inputenc}
\usepackage{pgfplots}
\DeclareUnicodeCharacter{2212}{−}
\usepgfplotslibrary{groupplots,dateplot,patchplots}
\usetikzlibrary{patterns,shapes.arrows}
\pgfplotsset{compat=newest}

\pgfplotscreateplotcyclelist{cbw}{%
    blue,           every mark/.append style={                     fill=blue!80!black   }, mark=o\\%
    green!50!black, dashed, every mark/.append style={                     fill=green!30!black  }, mark=x\\%
    brown!60!black,densely dotted, every mark/.append style={                     fill=brown!80!black  }, mark=triangle*\\%
    black,                                                                                 mark=star\\%
    red,            dashed,every mark/.append style={solid,fill=red!80!black    }, mark=*\\%
                    densely dashed,every mark/.append style={solid,fill=brown!80!black  }, mark=square*\\%
    black,          densely dashed,every mark/.append style={solid,fill=gray            }, mark=otimes*\\%
    blue,           densely dashed,every mark/.append style={solid                      }, mark=star\\%
    red,            densely dashed,every mark/.append style={solid,fill=red!80!black    }, mark=diamond*\\%
}
\usepackage{tikz}

\usetikzlibrary{calc,trees,positioning,arrows,chains,shapes.geometric,%
    decorations.pathreplacing,decorations.pathmorphing,shapes,%
    matrix,shapes.symbols}

\tikzstyle{line} = [draw, -, thick]
\tikzstyle{nodraw} = [draw, fill, circle, minimum width=0pt, inner sep=0pt]
\tikzstyle{box} = [line, rectangle, rounded corners, text centered]

\tikzset{
>=stealth',
  punktchain/.style={
    rectangle, 
    rounded corners, 
    draw=black, very thick,
    text width=2em, 
    minimum height=3em, 
    text centered, 
    on chain},
  line/.style={draw, thick, <-},
  element/.style={
    tape,
    top color=white,
    bottom color=blue!50!black!60!,
    minimum width=1em,
    draw=blue!40!black!90, very thick,
    text width=2em, 
    minimum height=3.5em, 
    text centered, 
    on chain},
  every join/.style={->, thick,shorten >=1pt},
  decoration={brace},
  tuborg/.style={decorate},
  tubnode/.style={midway, right=2pt},
}

\usepackage{environ}
\makeatletter
\newsavebox{\measure@tikzpicture}
\NewEnviron{scaletikzpicturetowidth}[1]{%
  \def\tikz@width{#1}%
  \begin{lrbox}{\measure@tikzpicture}%
  \BODY
  \end{lrbox}%
  \pgfmathparse{#1/\wd\measure@tikzpicture}%
  \BODY
}
\makeatother
\pgfplotsset{every tick label/.append style={font=\small}}
\usepackage{amssymb}
\usepackage{graphicx}
\usepackage{amsmath}
\usepackage{xcolor}
\usepackage{url}
\newcommand{\atol}[0]{{\em atol~}}

\begin{document}
\nolinenumbers
\title{Lossy Checkpoint Compression in Full Waveform Inversion: a case study with ZFPv0.5.5 and the Overthrust Model}


\Author[1]{Navjot}{Kukreja}
\Author[2]{Jan}{H\"uckelheim}
\Author[3]{Mathias}{Louboutin}
\Author[4]{John}{Washbourne}
\Author[5]{Paul H.J.}{Kelly}
\Author[1]{Gerard J.}{Gorman}
\affil[1]{Department of Earth Science and Engineering, Imperial College London }
\affil[2]{Argonne National Laboratory}
\affil[3]{Georgia Institute of Technology}
\affil[4]{Chevron Corporation}
\affil[5]{Department of Computing, Imperial College London}




\correspondence{Navjot Kukreja (n.kukreja@liverpool.ac.uk)}

\runningtitle{Lossy Checkpoint Compression in FWI}

\runningauthor{N. Kukreja et al.}

\received{}
\pubdiscuss{} 
\revised{}
\accepted{}
\published{}


\firstpage{1}

\maketitle

\begin{abstract}
This paper proposes a new method that combines checkpointing methods with error-controlled lossy compression for large-scale high-performance Full-Waveform Inversion (FWI), an inverse problem commonly used in geophysical exploration. This combination can significantly reduce data movement, allowing a reduction in run time as well as peak memory.

In the Exascale computing era, frequent data transfer (e.g., memory bandwidth, PCIe bandwidth for GPUs, or network) is the performance bottleneck rather than the peak FLOPS of the processing unit. 

Like many other adjoint-based optimization problems, FWI is costly in terms of the number of floating-point operations, large memory footprint during backpropagation, and data transfer overheads. Past work for adjoint methods has developed checkpointing methods that reduce the peak memory requirements during backpropagation at the cost of additional floating-point computations.

Combining this traditional checkpointing with error-controlled lossy compression, we explore the three-way tradeoff between memory, precision, and time to solution. We investigate how approximation errors introduced by lossy compression of the forward solution impact the objective function gradient and final inverted solution. Empirical results from these numerical experiments indicate that high lossy-compression rates (compression factors ranging up to 100) have a relatively minor impact on convergence rates and the quality of the final solution.

\keywords{Lossy compression, Full waveform inversion, checkpointing, memory}
\end{abstract}


\introduction  
Full-waveform inversion (FWI) is an adjoint-based optimization problem used in seismic imaging to infer the earth's subsurface structure and physical parameters \citep{virieux2009overview}. The compute and memory requirements for this and similar PDE-constrained optimization problems can readily push the world's top supercomputers to their limits. Table~\ref{tab:reqs} estimates the computational requirements of an FWI problem on the SEAM Model \citep{SEAM}. Although the grid-spacing and timestep interval depends on various problem-specific factors, we can do a back-of-the-envelope calculation to appreciate the scale of FWI. To estimate the number of operations per grid point, we use a variant of Equation~\ref{eqn:wave} called TTI \citep{tti-main}, which is commonly used today in commercial FWI. Such a problem would require almost 90 days of continuous execution at 1 PFLOP/s. The memory requirements for this problem are also prohibitively high. As can be seen in Table~\ref{tab:reqs}, the gradient computation step is responsible for this problem's high memory requirement, and the focus of this paper is to reduce that requirement.

The FWI algorithm is explained in more detail in Section~\ref{sec:fwi}. It is important to note that despite the similar terminology, the checkpointing we refer to in this paper is not done for resilience or failure recovery. This is the checkpointing from automatic-differentiation theory, with the objective of reducing the memory footprint of a large computation by trading recomputation for storage.

\begin{table}[htbp]
\centering
\begin{tabular}{|c|c|c|c|}
\hline
\textbf{Description} & \textbf{Number} & \textbf{Peak Memory} & \textbf{No. of Flops} \\
\hline
Single grid point (TTI) & 1 & 8 bytes & 6300 \\
\hline
Complete grid & $1000 \times 1000 \times 1000$ & 8GB & $6.3 \times 10^{12}$  \\
\hline
Forward propagation & 10000 & 24GB & $6.3 \times 10^{16}$   \\
\hline
Gradient Computation & 2 (FW+REV)$^1$ & 80TB & $1.26 \times 10^{17}$\\
\hline
Shots & 10000 & 80TB & $1.26 \times 10^{21}$\\
\hline
Optim. Iterations & 20 & 80TB & $2.52 \times 10^{22}$\\
\hline
\end{tabular}
\vspace{0.5em}
\caption{Estimated computational requirements of a Full-Waveform Inversion problem based on the SEAM model \protect{\citep{SEAM}}, a large scale industry standard geophysical model that is used to benchmark FWI. Note that real-world FWI problems are likely to be larger. $^1$A gradient computation involves a forward simulation followed by a reverse/adjoint computation. For simplicity we assume the same size of computation during the forward/adjoint pass.}
\label{tab:reqs}
\end{table}

\subsection{FWI and other similar problems}
FWI is similar to other inverse problems like brain-imaging \citep{guasch2020full}, shape optimization \citep{jameson1998optimum}, and even training a neural network. When training a neural network, the activations calculated when propagating forward along the network need to be stored in memory and used later during backpropagation. The size of the corresponding computation in a neural network depends on the depth of the network and, more importantly, the input size. We assume the input is an image for the purpose of this exposition. For typical input image sizes of less than $ 500 \times 500$ px, the computation per data point is relatively (to FWI) small, both in the number of operations and memory required. This is compensated by processing in \emph{mini-batches}, where multiple data points are processed at the same time. This batch dimension's size is usually adjusted to \emph{fill up} the target hardware to its capacity (and no more). This is the standard method of managing the memory requirements of a neural network training pipeline. However, for an input image that is large enough, or a network that is deep enough, it is seen that the input image, network weights, and network activations together require more memory than available on a single node, even for a single input image (\emph{batchsize} $=1$ ). We previously addressed this issue in the context of neural networks \citep{kukreja2019training}. In this paper we address the same issue for FWI. 

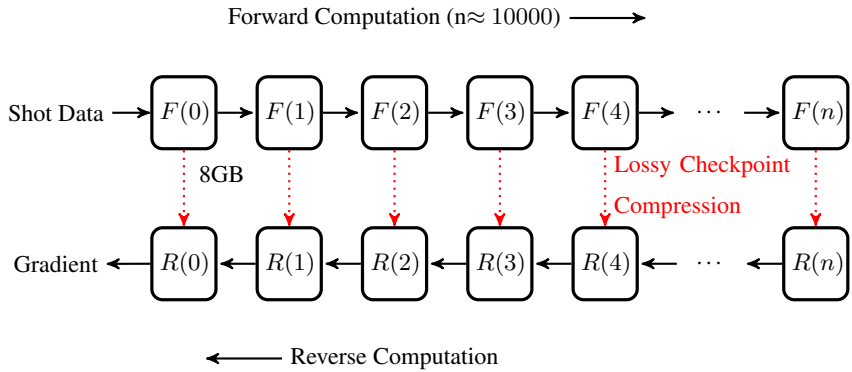
\begin{figure}[htbp]
    \centering
    \begin{tikzpicture}
  [node distance=.5cm,
  start chain=1 going right,start chain=2 going left]
    \node[] at (0,0) (shot) {Shot Data};
    \node[below=1.55cm of shot] (gradient) {Gradient};
     \node[punktchain, join, on chain=1, right= of shot] (t1) {$F(0)$};
     \node[punktchain, join, on chain=1] (t2)      {$F(1)$};
     \node[punktchain, join, on chain=1] (t3)      {$F(2)$};
     \node[punktchain, join, on chain=1] (t4) {$F(3)$};
     \node[punktchain, join, on chain=1] (t5) {$F(4)$};
     \node[punktchain, join, draw=none, on chain=1] (ellipsis1) {$\cdots$};
     \node[punktchain, join, on chain=1] (tn) {$F(n)$};

\node[punktchain, below=1cm of tn, on chain=2] (atn) {$R(n)$};
\node[punktchain, join, draw=none, on chain=2] (ellipsis2) {$\cdots$};
     \node[punktchain, join, on chain=2] (at5)      {$R(4)$};
     \node[punktchain, join, on chain=2] (at4) {$R(3)$};
     \node[punktchain, join, on chain=2] (at3) {$R(2)$};
     \node[punktchain, join, on chain=2] (at2) {$R(1)$};
     \node[punktchain, join, on chain=2] (at1) {$R(0)$};

\node[above=0.5cm of t3] (forward) {Forward Computation (n$\approx 10000$)};
\node[below=0.5cm of at3] (reverse) {Reverse Computation};

\node[right=1cm of forward] (hiddenforward) {};
\node[left=1cm of reverse] (hiddenreverse) {};

\draw[->, thick,] (forward.east) -- (hiddenforward.west);
\draw[->, thick,] (reverse.west) -- (hiddenreverse.east);

\draw[->, thick,] (shot.east) -- (t1.west);
\draw[->, thick,] (at1.west) -- (gradient.east);

\draw[dotted,->, thick,red] (t1.south) |-+(0,-1em)-| node[right, black] {~8GB} (at1.north);
\draw[dotted,->, thick,red] (t2.south) |-+(0,-1em)-| (at2.north);
\draw[dotted,->, thick,red] (t3.south) |-+(0,-1em)-| (at3.north);
\draw[dotted,->, thick,red] (t4.south) |-+(0,-1em)-| (at4.north);
\draw[dotted,->, thick,red] (t5.south) |-+(0,-1em)-| node[auto, text width=4cm] {\\Lossy Checkpoint \\ Compression} (at5.north);
\draw[dotted,->, thick,red] (tn.south) |-+(0,-1em)-| (atn.north);
  \end{tikzpicture}
    \caption{An illustration of the approach presented in this paper. Checkpoints are compressed using lossy compression to combine lossy compression and the checkpoint-recompute strategies. }
    \label{fig:title}
\end{figure}

Many algorithmic optimizations/approximations are commonly applied to reduce the computational load from the numbers calculated in Table~\ref{tab:reqs}. These optimizations could either reduce the number of operations or the amount of memory required. In this paper, we shall focus on the high memory footprint of this problem. One standard approach is to save the field at only the boundaries and reconstruct the rest of the field from the boundaries to reduce the memory footprint. However, the applicability of this method is limited to time-reversible PDEs. In this work, we use the isotropic acoustic equation as the example (See Equation~\ref{eqn:wave}). Although this equation is time-reversible, many other variations used in practice are not. For this reason, we do not discuss this method in this paper.

A commonly used method to deal with the problem of this large memory footprint is domain-decomposition over MPI, where the computational domain is split into subdomains over multiple compute nodes to use their memory. The efficacy of this method depends on the ability to hide the MPI-communication overhead behind the computations within the subdomain. For effective communication-computation overlap, the subdomains should be big enough that the computations within the subdomain take at least as long as the MPI communication. This places a lower bound on subdomain size (and hence peak memory consumption per MPI rank) that is a function of the network interconnect - this lower bound might be too large for slow interconnects e.g. on cloud systems.

Some iterative frequency domain methods, e.g. \citet{knibbe2014closing}, can alleviate the memory limit but are not competitive with time-domain methods in the total time to solution. 

Hybrid methods that combine time-domain methods, as well as frequency-domain methods, have also been tried \citep{OTFFFT}. However, this approach can be challenging because the application code must decide the user's discrete set of frequencies to achieve a target accuracy.

In the following subsections, we discuss three techniques that are commonly used to alleviate this memory pressure - namely numerical approximations, checkpointing, and data compression. The common element in these techniques is that all three solve the problem of high memory requirement by increasing the operational intensity of the computation - doing more computations per byte transferred from memory. With the gap between memory and computational speeds growing wider as we move into the exaflop era, we expect to use such techniques to increase moving forward. 

\subsection{Approximate methods}

There has been some recent work on alternate floating-point representations \citep{chatelain2019automatic}, although we are not aware of this technique being applied to FWI. Within FWI, many approximate methods exist, including \emph{On-the-fly Fourier transforms} \citep{OTFFFT}. However, it is not clear whether this method can provide fine-tuned bounds on the solution's accuracy. In contrast, other completely frequency-domain formulations can provide clearer bounds \citep{sloppiness}, but as previously discussed, this comes at the cost of a much higher computational complexity. In this paper, we restrict ourselves to time-domain approaches only. 

Another approximation commonly applied to reduce the memory pressure in FWI in the time domain is subsampling. Here, the timestep-rate of the gradient computation (See equation~\ref{eq:Grad}) is decoupled from the timestep-rate of the adjoint wavefield computation, with one gradient timestep for every $n$ adjoint steps. This reduces the memory footprint by a factor of $n$, since only one-in-$n$ values of  the forward wavefield need to be stored. The Nyquist theorem is commonly cited as the justification for this sort of subsampling. However, the Nyquist theorem only provides a lower bound on the error - it is unclear whether an upper bound on the error has been established on this method. Although more thorough empirical measurements of the errors induced in subsampling have been done before \citep{louboutin2015time}, we do a brief empirical study in Section~\ref{sec:subsampling} as a baseline to compare the error with our method.

\subsection{Checkpointing}
Instead of storing the wavefield at every timestep during the forward computation, it is possible to store it at a subset of the timesteps only. During the following computation that proceeds in a reverse order to calculate the gradient, if the forward wavefield is required at a timestep that was not stored, it can be recovered by restarting the forward computation from the last available timestep. This is commonly known as checkpointing. Algorithms have been developed to define the optimal checkpointing schedule involving forward, store, backward, load, and recompute events under different assumptions \citep{revolve,wang2009,aupy2017periodicity}. This technique has also been applied to FWI-like computations \citep{symes2007reverse}.

\begin{figure}
    \centering
    \includegraphics[width=\textwidth]{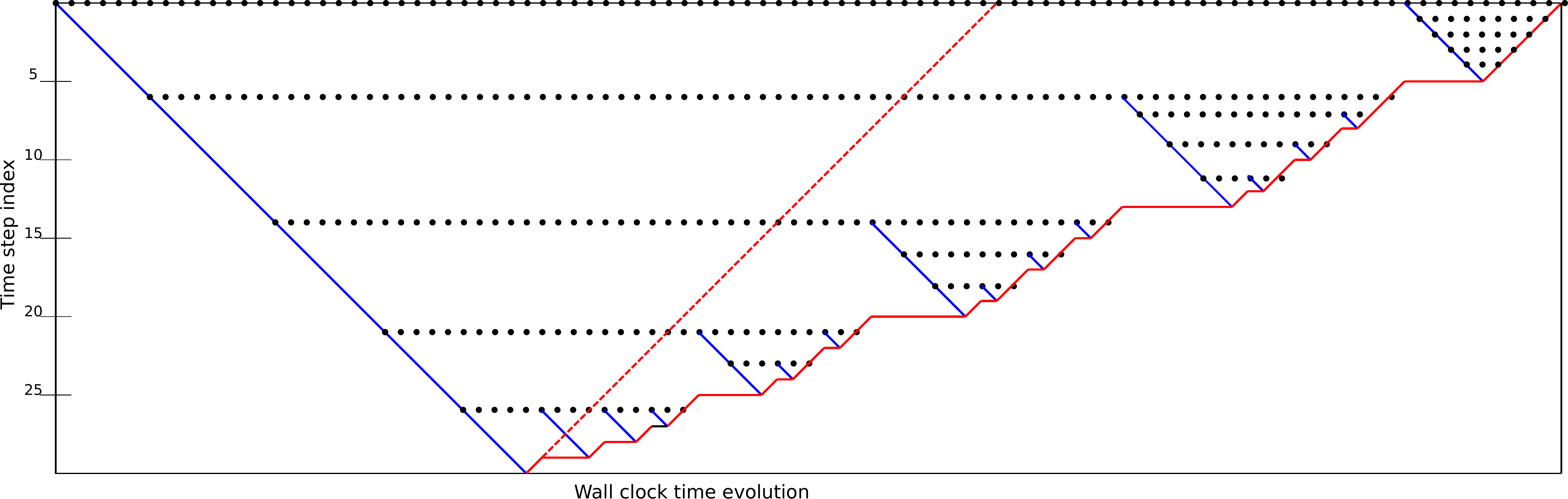}
    \caption{Schematic of the checkpointing strategy. Wall-clock time is on the horizontal axis, while the vertical axis represents simulation time. The blue line represents forward computation. The dotted red line represents how the reverse computation would have proceeded after the forward computation, had there been enough memory to store all the necessary checkpoints. Checkpoints are shown as the black dots. The reverse computation under the checkpointing strategy is shown as the solid red line. It can be seen that the reverse computation proceeds only where the results of the forward computation are available. When not available, the forward computation is restarted from the last available checkpoint to recompute the results of the forward.}
    \label{fig:checkpointing}
\end{figure}
In previous work, we introduced the open-source software pyRevolve, a Python module that can automatically manage the checkpointing strategies under different scenarios with minimal modification to the computations code \citep{kukreja2018high}. For this work, we extended pyRevolve to integrate lossy compression. 

The most significant advantage of checkpointing is that the numerical result remains unchanged by applying this technique. Note that we will shortly combine this technique with lossy compression which might introduce an error, but checkpointing alone is expected to maintain bitwise equivalence. Another advantage is that the increase in run time incurred by the recomputation is predictable. 

\subsection{Data Compression}

Compression or bit-rate reduction is a concept originally from signal processing. It involves representing information in fewer bits than the original representation. Since there is usually some computation required to go from one representation to another, compression can be seen as a memory-compute tradeoff. 

Perhaps the most commonly known and used compression algorithm is ZLib (from GZip) \citep{deutsch1996zlib}. TZLib is a lossless compression algorithm, i.e., the data recovered after compressing-decompressing is an exact replica of the original data before compression. Although ZLib is targeted at text data, which is one-dimensional and often has predictable repetition, other lossless compression algorithms are designed for other kinds of data. One example is FPZIP \citep{lindstrom2017fpzip}, which is a lossless compression algorithm for multidimensional floating-point data.

For floating-point data, another possibility is lossy compression, where the compressed-decompressed data is not exactly the same as the original data, but a close approximation. The precision of this approximation is often set by the user of the compression algorithm. Two popular algorithms in this class are SZ \citep{di2016fast} and ZFP \citep{lindstrom2014fixed}. 

Compression has often been used to reduce the memory footprint of adjoint computations in the past, including \citet{weiser2012state,boehm2016wavefield,marin2016large}. However, all these studies use hand-rolled compression algorithms specific to the corresponding task - \citet{weiser2012state} focusses on parabolic equations, \citet{boehm2016wavefield} focusses on wave propagation like us, and \citet{marin2016large} focusses on fluid flow. All three use their own lossy compression algorithm to compress the entire time history, and look at checkpointing as an alternative to lossy compression. In this paper we use a more general floating-point compression algorithm - ZFP. Since this compressor has been extensively used and studied across different domains and has implementations for various hardware platforms - this lends a sense of trust in this compressor, increasing the relevance of our work. None of the previously mentioned studies combine compression and checkpointing, as we do here.

\citet{cyr2015towards} performs numerical experiments to study the propagation of errors through an adjoint problem using compression methods like PCA. However, they do not consider the combination of checkpointing and compression in a single strategy. 

Floating-point can be seen as a \emph{compressed representation} that is not entirely precise. However, the errors introduced by the floating-point representation are already accounted for in the standard numerical analysis as noise. The errors introduced by ZFP's compression of fields are more subtle since the compression loss is pattern sensitive. Hence we tackle it empirically here. 

\begin{figure}[htbp]
    \centering
    \includegraphics[width=0.7\textwidth]{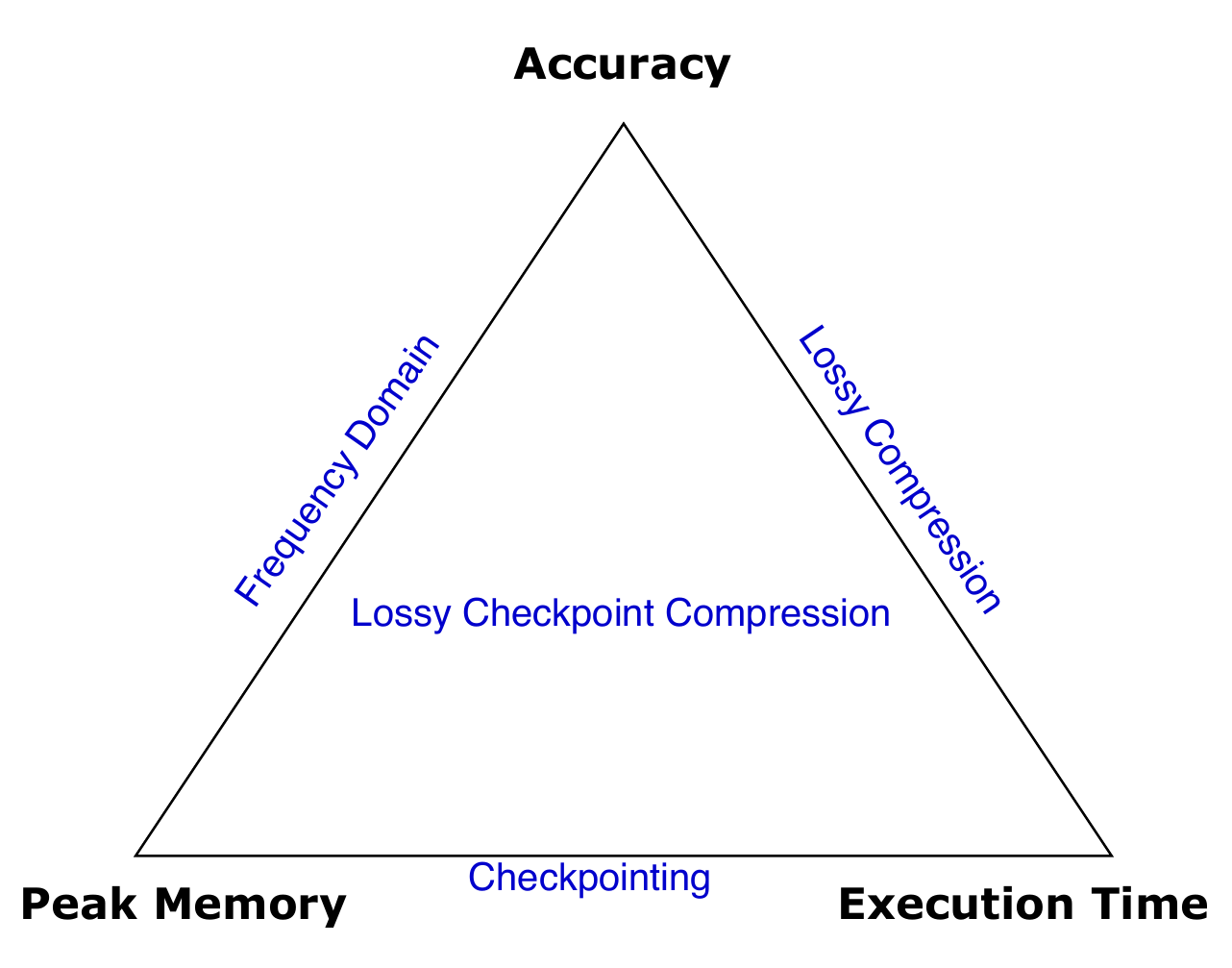}
    \caption{Schematic of the three-way tradeoff presented in this paper. With the use of checkpointing, it was possible to trade off memory and execution time (the horizontal line).  With the use of compression alone, it was possible to trade off memory and accuracy. The combined approach presented in this work provides a novel three-way tradeoff. }
    \label{fig:tradeoff}
\end{figure}

\subsection{Contributions}
The last few sections discussed some existing methods that allow trade-offs that are useful in solving FWI on limited resources. While checkpointing allows a trade-off between computational time and memory, compression allows a trade-off between memory and accuracy. This work combines these three approaches into one three-way trade-off.

In previous work \citep{kukreja2019combining}, we have shown that it is possible to accelerate generic adjoint-based computations (of which FWI is a  subset), by using lossy compression on the checkpoints. For a given checkpoint absolute error tolerance (\atol), compression may or may not accelerate the computation. The performance model from \citet{kukreja2019combining} helps us answer this question a priori, i.e., without running any computations. 

In this work, we evaluate this method on the specific problem of FWI, specifically the solver convergence and accuracy. 

To this end, we conduct an empirical study of:
\begin{enumerate}
  \item Propagation of errors when starting from a lossy checkpoint.
  \item Effect of checkpoint errors on the gradient computation.
  \item Effect of decimation/subsampling on the gradient computation.
  \item Accumulation of errors through the stacking of multiple shots.
  \item Effect of the lossy gradient on the convergence of FWI.
\end{enumerate}

The rest of the paper is organized as follows. Section~\ref{sec:fwi} gives an overview of FWI. This is followed by a description of our experimental setup in Section~\ref{sec:experiment}. Next, Section~\ref{sec:results} discusses the results, followed by our conclusions.

\section{Full Waveform Inversion}
\label{sec:fwi}
FWI is designed to numerically simulate a seismic survey experiment and invert for the earth parameters that best explain the observations. In the physical experiment, a ship sends an acoustic impulse through the water by triggering an explosion. The waves created as a result of this impulse travel through the water into the earth's subsurface. The reflections and turning components of these waves are recorded by an array of receivers being dragged in tow by the ship. A recording of one signal sent and the corresponding signals received at each of the receiver locations is called a shot. A single experiment typically consists of $~10000$ shots. 

Having recorded this collection of data ($\mathbf{d}_{\text{obs}}$), the next step is the numerical simulation. This starts with a wave equation. Many equations exist that can describe the propagation of a sound wave through a medium - the choice is usually a trade-off between accuracy and computational complexity. We mention here the simplest such equation, that describes isotropic acoustic wave propagation:
\begin{equation}
\mathbf{m}(x)\frac{\partial^2 \mathbf{u}(t, x)}{\partial t^2} - \nabla^2 \mathbf{u}(t, x) = \mathbf{q}_s(t, x),
\label{eqn:wave}
\end{equation}
where $\mathbf{m}(x) = \frac{1}{\mathbf{c}^2(x)}$ is the squared slowness, $\mathbf{c}(x)$ the spatially
dependent speed of sound, $\mathbf{u}(t, x)$ is the pressure wavefield, $\nabla^2 \mathbf{u}(t, x)$ denotes the laplacian of the wavefield and $\mathbf{q}_s(t, x)$ is a source term. Solving Equation~\ref{eqn:wave} for a given $\mathbf{m}$ and $\mathbf{q}_s$ can give us the \emph{simulated} signal that \emph{would be} received at the receivers. Specifically, the \emph{simulated data} can be written as:
\begin{equation}
\label{eq:rec_data}
\mathbf{d}_{\text{sim}} = \mathbf{P}_r \mathbf{u} = \mathbf{P}_r \mathbf{A}(\mathbf{m})^{-1} \mathbf{P}_s^\top \mathbf{q}_s
\end{equation}
where $\mathbf{P}_r$ is the measurement operator that restricts the full wavefield to the receivers locations, $\mathbf{A}(\mathbf{m})$ is the linear operator that is the discretization of the operator corresponding to Equation~\ref{eqn:wave}, and $\mathbf{P}_s$ is a linear operator that injects a localized source ($\mathbf{q}_s$) into the computational grid.

Using this, it is possible to set up an optimization problem that aims to find the value of $\mathbf{m}$ that minimizes the difference between the simulated signal ($\mathbf{d}_{\text{sim}}$) and the observed signal ($\mathbf{d}_{\text{obs}}$):

\begin{equation}
\operatorname*{argmin}_{\mathbf{m}} \Phi_s(\mathbf{m}) = \frac{1}{2} \left\lVert \mathbf{d}_{\text{sim}} - \mathbf{d}_{\text{obs}} \right\rVert_2^2.
\end{equation}

This objective function $\Phi_s(\mathbf{m})$ can be minimized using a gradient descent method. The gradient can be computed as follows:
\begin{equation}
\label{eq:Grad}
 \nabla\Phi_s(\mathbf{m})=\sum_{t =1}^{n_t}\mathbf{u}[\mathbf{t}] \mathbf{v}_{tt}[\mathbf{t}] =\mathbf{J}^T\delta\mathbf{d}%
\end{equation}
where $\mathbf{u}[\mathbf{t}] $ is the wavefield from Equation~\ref{eqn:wave} and $\mathbf{v}_{tt}[\mathbf{t}]$ is the second-derivative of the adjoint field \citep{tarantola1984inversion}. The adjoint field is computed by solving an adjoint equation \emph{backwards} in time. The appropriate adjoint equation is a result of the choice of the forward equation. In this example, we chose the acoustic isotropic equation (Equation~\ref{eqn:wave}), which is self-adjoint. However, it is not always trivial to derive the adjoint equation corresponding to a chosen forward equation \citep{huckelheim2019automatic}. This adjoint computation can only be started once the forward computation (i.e. the one involving Equation~\ref{eqn:wave}) is complete. Commonly, this is done by storing the intermediate values of $\mathbf{u}$ during the forward computation, then starting the adjoint computation to get values of $\mathbf{v}$, and using that and the previously calculated $\mathbf{u}$ to directly calculate $\nabla\Phi_s(\mathbf{m})$ in the same loop. This need to store the intermediate values of $\mathbf{u}$ during the forward computation is the source of the high memory footprint of this method. 

The computation described in the previous paragraph is for a single shot and must be repeated for every shot, and the final gradient is calculated by averaging the gradients calculated for the individual shots. This is repeated for every iteration of the minimization. This entire minimization problem is one step of a multi-grid method that starts by inverting only the low frequency components on a coarse grid, and adding higher frequency components which require finer grids over successive inversions. 
\section{Experimental setup}
\label{sec:experiment}
\begin{description}

\item[Reference Problem] We use Devito \citep{kukreja2016devito,luporini2018architecture, louboutin2019devito} to build an acoustic wave propagation experiment. The velocity model was initialized using the SEG Overthrust model. This velocity model was then smoothed using a Gaussian function to simulate a starting guess for a complete FWI problem. The original domain was surrounded by a 40 point deep absorbing boundary layer. This led to a total of $287 \times 881 \times 881$ grid points. This was run for $4000$ms with a step of $1.75$ms, making 2286 timesteps. The spatial domain was discretized on a grid with a grid spacing of 20m, and the discretization was 16th-order in space and second-order in time. We used 80 shots for our experiments with the sources placed along the x-dimension, spaced equally and just under the water surface. The shots were generated by modeling a Ricker source of peak frequency 8Hz. Following the method outlined in \citet{peters2019projection}, we avoid inverse crime by generating the shots using a variation of Equation~\ref{eqn:wave} that includes density, while using Equation~\ref{eqn:wave} for inversion. The gradient was scaled by dividing by the norm of the original gradient in the first iteration. This problem solved in double precision is what we shall refer to as the \emph{reference problem} in the rest of this paper. Note that this reference solution itself has many sources of error, including floating-point arithmetic and the discretization itself. 

\item[Evolution of compressibility] We attempt to compress every timestep of the reference problem using the same compression setting and report on the achieved compression factor as a function of the timestep. 

\item[Direct compression] Based on the previous experiment, we choose a reference wavefield and compress it directly using a variety of compression settings. In this experiment, we report the errors comparing the lossy wavefield and the true reference wavefield. 

\item[Forward propagation] In this experiment, we run the forward simulation for a few timesteps (about half the reference problem) and store it as a checkpoint. We then compress and decompress this through the lossy compression algorithm, getting two checkpoints - a reference checkpoint and a lossy checkpoint. We restart the simulation from each of these checkpoints and compare the two simulations' states and report on differences. 

\item[Gradient Computation] In this experiment, we do the complete gradient computation, as shown in Figure~\ref{fig:title} - once for the reference problem and a few different lossy settings. We report on the differences between these to show the propagation of errors. 

\item[Stacking] In this experiment, we collate the gradient computed on multiple shots, i.e., all ten shots, and report the difference between the reference problem and the compressed version for this step. 

\item[Convergence] In practice, FWI is run for only a few iterations at a time as a fine-tuning step interspersed with other imaging steps. Here we run a fixed number of FWI iterations (30) to make it easier to compare different experiments. To make this a practical test problem, we extract a 2D slice from the original 3D velocity model and run a 2D FWI instead of 3D. We compare the convergence trajectory with the reference problem and report. 

\item[Subsampling] As a comparison baseline, we also use subsampling to reduce the memory footprint as a separate experiment and track the errors. The method is set up so that the forward and adjoint computations continue at the same time stepping as the reference problem above. However, the gradient computation is now not done at the same rate - it is reduced by a factor \emph{f}. We plot results for varying \emph{f}.
\end{description}

\subsection{Error metrics}
In this work, we only ever compress the forward wavefield ($u$ from Section~\ref{sec:fwi}) using lossy compression. Let $F(i, j, k)$ be the original field (i.e. before any compression/loss), and $G(i, j, k)$ be the field recovered after lossy compression of $F(i, j, k)$, followed by decompression. 
We report errors using the following metrics:
\begin{description}
\item[PSNR]: Peak Signal to Noise Ratio, we define this as:
\begin{equation}
    \text{PSNR (dB)} = 10 \text{log}_{10}\frac{R^2}{\text{MSE}}, 
\end{equation}
where $R$ is the range of values in the field to be compressed, and $\text{MSE}$ is the \emph{mean squared error} between the reference and the lossy field. More precisely,  
\begin{equation}
    \text{MSE} = \frac{1}{mnp}\sum\limits_{k=0}^{p} \sum\limits_{j=0}^{n} \sum\limits_{i=0}^{m} [F(i, j, k) - G(i, j, k)]^2, 
\end{equation}
and $R=\text{max}(F(i, j, k)) - \text{min}(F(i, j, k))$.

\item[Angle]: We treat $F(i, j, k)$ and $G(i, j, k)$ as vectors and calculate the angle between them as follows:
\begin{equation}
    cos \theta = \frac{\overrightarrow{F} \cdot \overrightarrow{G}}{\left\lVert\overrightarrow{F}\right\rVert \cdot \left\lVert\overrightarrow{G}\right\rVert}
\end{equation}

\item[Error Norms]: We also report some errors by defining the error vector $E(i, j, k)$, as $F(i, j, k) - G(i, j, k)$, and reporting $L_2$ and $L_\infty$ norms of this vector.
\end{description}

\section{Results}
\label{sec:results}

\subsection{Evolution of compressibility}
\begin{figure}[htbp]
    \centering
    \begin{scaletikzpicturetowidth}{\textwidth}
    \input{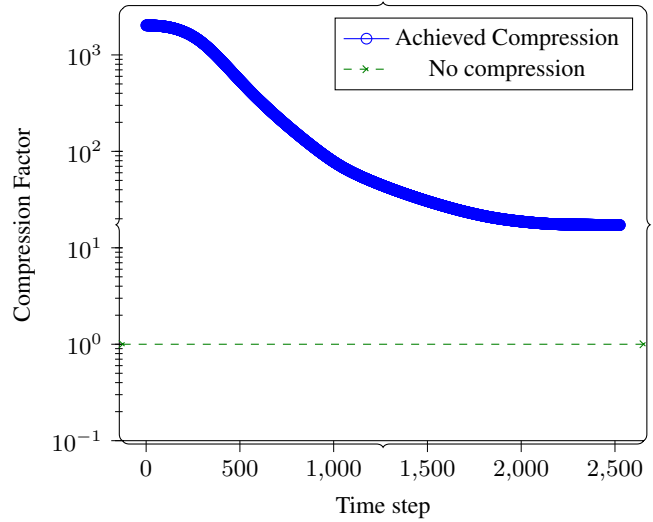}
    \end{scaletikzpicturetowidth}
    \caption{Evolution of compressibility through the simulation. We tried to compress every time step of the reference problem using an absolute error tolerance (\atol) setting of $10^{-4}$. The Compression factor achieved is plotted here as a function of the timestep number. Higher is more compression. Dotted line represents \emph{no compression}. We can see that the first few timesteps are compressible to 1000x - since they are mostly zeros. The achievable compression factor drops as the wave propagates through the domain and seems to stabilize to ~20x towards the end. We pick the last time step as the reference field for further experiments.}
    \label{fig:compression_ratios}
\end{figure}
To understand the evolution of compressibility, we tried to compress each and every timestep of a simulation to observe the evolution of compressibility through the simulation. 
This is shown in Figure~\ref{fig:compression_ratios}. It can be seen that in the beginning the field is highly compressible since it 
consists of mostly zeros. The compressibility is worst towards the end of the simulation when the wave has reached most of the domain.

\begin{figure}[htbp]
    \centering
    \includegraphics[width=\textwidth]{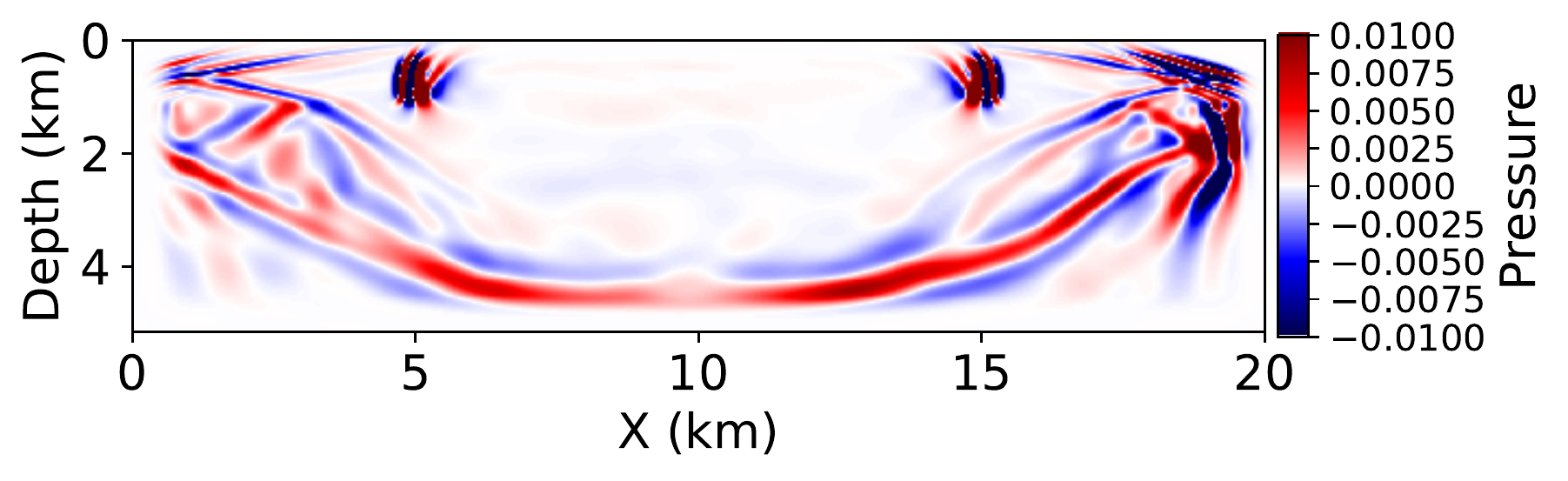}
    \caption{A 2D slice of the last time step of the reference solution. The wave has spread through most of the domain.}
    \label{fig:uncompressed}
\end{figure}
Therefore we pick the last timestep as the reference for further experiments. A 2D cross section of this snapshot is shown in Figure \ref{fig:uncompressed}.

\subsection{Direct compression}
\begin{figure}[htbp]
\centering
    \begin{scaletikzpicturetowidth}{0.4\textwidth}
\begin{tikzpicture}

\begin{axis}[
cycle list name=cbw, 
log basis x={10},
log basis y={10},
minor ytick={2,3,4,5,6,7,8,9,20,30,40,50,60,70,80,90,200,300,400,500,600,700,800,900,2000,3000,4000,5000,6000,7000,8000,9000,20000,30000,40000,50000,60000,70000,80000,90000},
,
tick align=outside,
tick pos=left,
x grid style={white!69.0196078431373!black},
xlabel={atol},
xmin=1.58489319246111e-17, xmax=6.30957344480193,
xmode=log,
xtick style={color=black},
xtick={1e-19,1e-17,1e-15,1e-13,1e-11,1e-09,1e-07,1e-05,0.001,0.1,10,1000},
xticklabels={\(\displaystyle {10^{-19}}\),\(\displaystyle {10^{-17}}\),\(\displaystyle {10^{-15}}\),\(\displaystyle {10^{-13}}\),\(\displaystyle {10^{-11}}\),\(\displaystyle {10^{-9}}\),\(\displaystyle {10^{-7}}\),\(\displaystyle {10^{-5}}\),\(\displaystyle {10^{-3}}\),\(\displaystyle {10^{-1}}\),\(\displaystyle {10^{1}}\),\(\displaystyle {10^{3}}\)},
y grid style={white!69.0196078431373!black},
ylabel={Peak Signal to Noise Ratio},
ymin=51.7929689100011, ymax=409.86155731391,
ymode=log,
ytick style={color=black},
ytick={1,10,100,200, 300, 400},
yticklabels={\(\displaystyle {10^{0}}\),\(\displaystyle {10^{1}}\),\(\displaystyle {100}\),\(\displaystyle {200}\),\(\displaystyle {300}\), ,\(\displaystyle {400}\)},
cycle list name=cbw,
scale only axis
]
\addplot
table {%
1 56.8991321054638
0.1 72.2931311796383
0.01 86.3998219384503
0.001 101.562815181703
0.0001 122.75747781663
1e-05 139.174237298576
1e-06 155.132859316986
1e-07 178.275416018082
1e-08 196.171336393158
1e-09 214.147998349171
1e-10 238.732105979528
1e-11 258.421158387981
1e-12 279.143787336361
1e-13 306.933317520216
1e-14 327.564521311696
1e-15 347.604920880084
1e-16 373.080328466478
};
\end{axis}

\end{tikzpicture}
    \end{scaletikzpicturetowidth}
   \caption{Direct compression: We compress the wave field at the last time step of the reference solution using different \atol  settings and report the Peak Signal to noise ratio (PSNR) achieved. Higher PSNR is lower error. The PSNR is very high for low absolute error tolerance (\atol) and drops predictably as \atol  is increased. See figures~\ref{fig:direct_error_Linf}~and~\ref{fig:direct_error_L1} for more metrics on this comparison.}\label{fig:direct_error_psnr}
\end{figure}
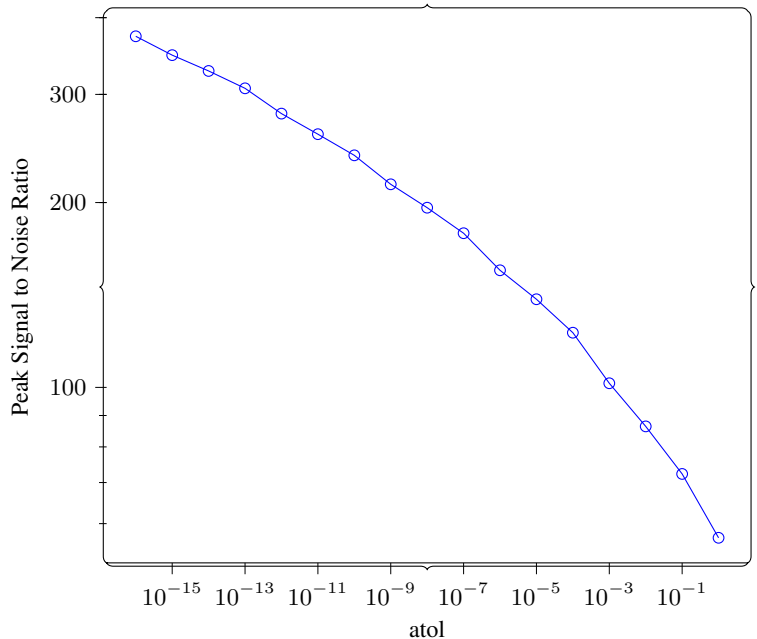

To understand the direct effects of compression, we compressed the reference wavefield using a variety of absolute tolerance (\atol) settings and observed the errors incurred as a function of \atol. The error is a tensor of the same shape as the original field and results from subtracting the reference field and the lossy field. Figure~\ref{fig:direct_error_psnr} shows the Peak Signal-to-Noise Ratio achieved for each \atol setting. Figure~\ref{fig:direct_error_Linf} in the appendix shows  some additional norms for this error tensor. 

\subsection{Forward propagation}

Next, we ran the simulation for 550 steps and compressed the field's final state after this time. We then restarted the simulation from step 550, comparing the progression of the simulation restarted from the lossy checkpoint vs. a \emph{reference} simulation that was started from the original checkpoint. We run this test for another 550 timesteps. 

\begin{figure}[htbp]
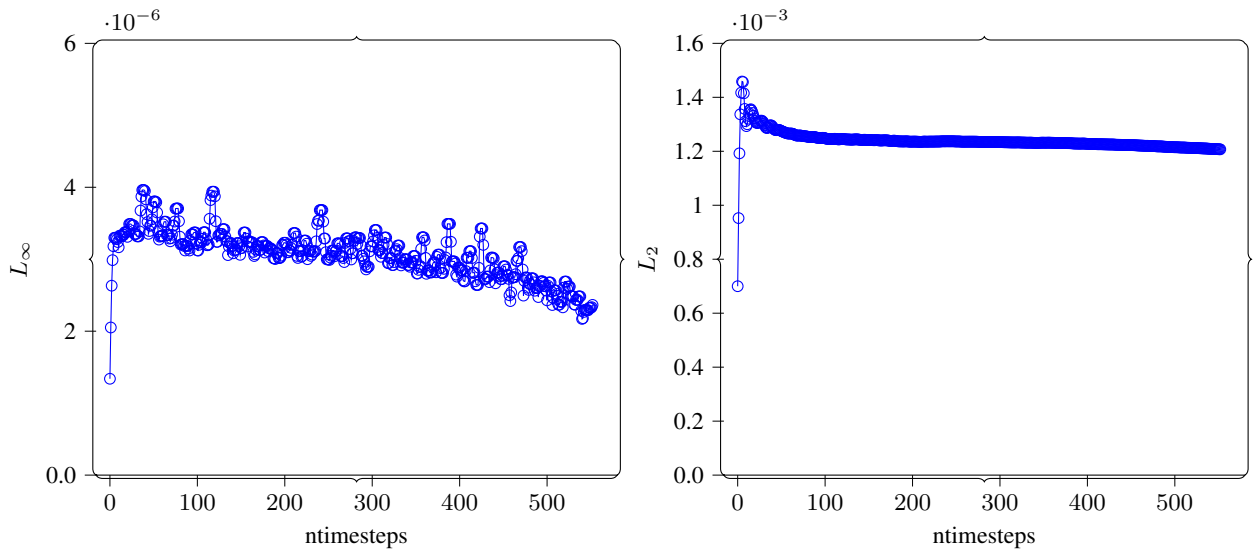

\centering
   \begin{scaletikzpicturetowidth}{0.4\textwidth}
        \input{figures/forward_Linf}\input{figures/forward_L2}
   \end{scaletikzpicturetowidth}
   \caption{Forward propagation: We stop the simulation after about 500 timesteps. We then compress the state of the wavefield at this point using \atol $=10^{-6}$. We then continue the simulation from the lossy checkpoint and compare with the reference version. Here we report the $L_\infty$ and $L_2$ norms of the error between the wavefields of these two versions as a function of the number of timesteps evolved from this lossy checkpoint. Both norms grow sharply for the first few timesteps before settling into a decreasing trend.}
   \label{forward_error}
\end{figure}

Figure~\ref{forward_error} shows the evolution of $L_\infty$ and $L_2$ norms as a function of the number of timesteps evolved. Both,  $L_\infty$ norm and $L_2$ norm grow sharply for the first few timesteps before stabilising into a downward trend. This tells us that the numerical method is robust to the error induced by lossy compression and the error does not appear to be forcing the system to a different solution. 

\subsection{Gradient computation}

Next, we measured the error in the gradient computation as a function of \atol, assuming the same compression settings are used for all checkpoints. 

\begin{figure}[htbp]
\centering
\begin{tikzpicture}

\begin{axis}[
cycle list name=cbw, 
log basis x={10},
tick align=outside,
tick pos=left,
x grid style={white!69.0196078431373!black},
xlabel={atol},
xmin=1.58489319246111e-17, xmax=6.30957344480193,
xmode=log,
xtick style={color=black},
xtick={1e-19,1e-17,1e-15,1e-13,1e-11,1e-09,1e-07,1e-05,0.001,0.1,10,1000},
xticklabels={\(\displaystyle {10^{-19}}\),\(\displaystyle {10^{-17}}\),\(\displaystyle {10^{-15}}\),\(\displaystyle {10^{-13}}\),\(\displaystyle {10^{-11}}\),\(\displaystyle {10^{-9}}\),\(\displaystyle {10^{-7}}\),\(\displaystyle {10^{-5}}\),\(\displaystyle {10^{-3}}\),\(\displaystyle {10^{-1}}\),\(\displaystyle {10^{1}}\),\(\displaystyle {10^{3}}\)},
y grid style={white!69.0196078431373!black},
ylabel={Peak Signal to Noise Ratio},
ymin=77.7775479355, ymax=196.9984755745,
ytick style={color=black}
]
\addplot
table {%
1 83.19668101
0.1 105.21503
0.01 122.2804414
0.001 139.7136278
0.0001 163.0279782
1e-05 177.5558469
1e-06 189.1345985
1e-07 191.5507928
1e-08 191.5753049
1e-09 191.579232
1e-10 191.5793425
1e-11 191.5793405
1e-12 191.5793406
1e-13 191.5793406
1e-14 191.5793406
1e-15 191.5793406
1e-16 191.5793406
};
\end{axis}

\end{tikzpicture}

   \caption{Gradient computation: In this experiment we carry out the full forward-reverse computation to get a gradient for a single shot, while compressing the checkpoints at different \atol settings. This plot shows the PSNR of true vs lossy gradient as a function of \atol on the lossy checkpoints. We can see that the PSNR remains unchanged until about \atol $=10^{-6}$ and is very high even at very high values of \atol.}\label{gradient_error_psnr}
\end{figure}
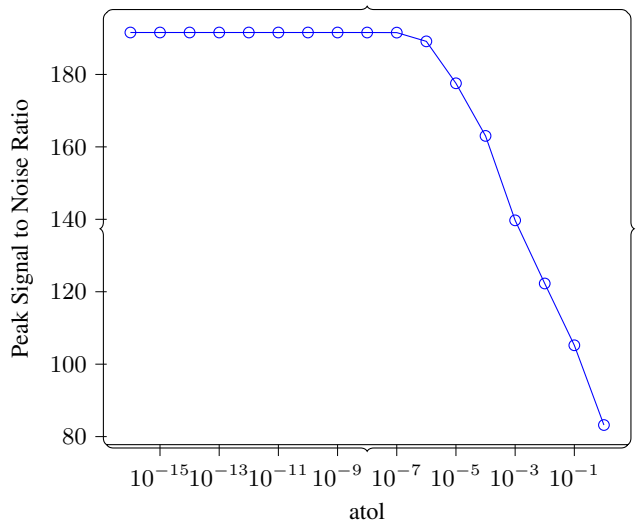

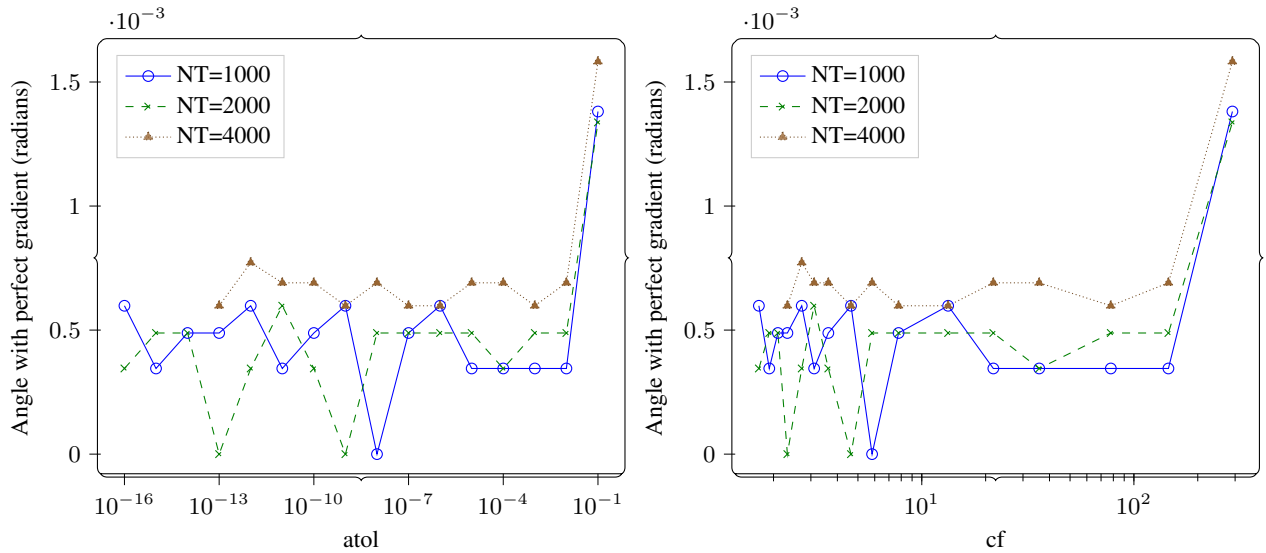
\begin{figure}[htbp]
\centering
   \begin{scaletikzpicturetowidth}{0.5\textwidth}
\begin{tikzpicture}

\begin{axis}[
cycle list name=cbw, 
legend cell align={left},
legend style={fill opacity=0.8, draw opacity=1, text opacity=1, at={(0.03,0.97)}, anchor=north west, draw=white!80!black},
log basis x={10},
tick align=outside,
tick pos=left,
x grid style={white!69.0196078431373!black},
xlabel={atol},
xmin=1.77827941003892e-17, xmax=0.562341325190349,
xmode=log,
xtick style={color=black},
y grid style={white!69.0196078431373!black},
ylabel={Angle with perfect gradient (radians)},
ymin=-7.9110612480493e-05, ymax=0.00166132286209035,
ytick style={color=black}
]
\addplot 
table {%
1e-16 0.000598019965645366
1e-15 0.000345266984716204
1e-14 0.000488281254850639
1e-13 0.000488281254850639
1e-12 0.000598019965645366
1e-11 0.000345266984716204
1e-10 0.000488281254850639
1e-09 0.000598019965645366
1e-08 0
1e-07 0.000488281254850639
1e-06 0.000598019965645366
1e-05 0.000345266984716204
0.0001 0.000345266984716204
0.001 0.000345266984716204
0.01 0.000345266984716204
0.1 0.00138106804176242
};
\addlegendentry{NT=1000}
\addplot
table {%
1e-16 0.000345266984716204
1e-15 0.000488281254850639
1e-14 0.000488281254850639
1e-13 0
1e-12 0.000345266984716204
1e-11 0.000598019965645366
1e-10 0.000345266984716204
1e-09 0
1e-08 0.000488281254850639
1e-07 0.000488281254850639
1e-06 0.000488281254850639
1e-05 0.000488281254850639
0.0001 0.000345266984716204
0.001 0.000488281254850639
0.01 0.000488281254850639
0.1 0.00133721337478927
};
\addlegendentry{NT=2000}
\addplot
table {%
1e-13 0.000598019965645366
1e-12 0.000772040463550879
1e-11 0.000690533979722166
1e-10 0.000690533979722166
1e-09 0.000598019965645366
1e-08 0.000690533979722166
1e-07 0.000598019965645366
1e-06 0.000598019965645366
1e-05 0.000690533979722166
0.0001 0.000690533979722166
0.001 0.000598019965645366
0.01 0.000690533979722166
0.1 0.00158221224960986
};
\addlegendentry{NT=4000}
\end{axis}

\end{tikzpicture}
\begin{tikzpicture}

\begin{axis}[
cycle list name=cbw, 
legend cell align={left},
legend style={fill opacity=0.8, draw opacity=1, text opacity=1, at={(0.03,0.97)}, anchor=north west, draw=white!80!black},
log basis x={10},
tick align=outside,
tick pos=left,
x grid style={white!69.0196078431373!black},
xlabel={cf},
xmin=1.31957497853861, xmax=376.767944506662,
xmode=log,
xtick style={color=black},
y grid style={white!69.0196078431373!black},
ylabel={Angle with perfect gradient (radians)},
ymin=-7.9110612480493e-05, ymax=0.00166132286209035,
ytick style={color=black}
]
\addplot
table {%
1.706294796 0.000598019965645366
1.909927969 0.000345266984716204
2.097305367 0.000488281254850639
2.324965044 0.000488281254850639
2.717075768 0.000598019965645366
3.108458024 0.000345266984716204
3.626176704 0.000488281254850639
4.638408358 0.000598019965645366
5.830579089 0
7.774222389 0.000488281254850639
13.312624 0.000598019965645366
21.74411405 0.000345266984716204
35.77491095 0.000345266984716204
77.85465534 0.000345266984716204
145.4687624 0.000345266984716204
291.3761171 0.00138106804176242
};
\addlegendentry{NT=1000}
\addplot
table {%
1.706294796 0.000345266984716204
1.909927969 0.000488281254850639
2.097305367 0.000488281254850639
2.324965044 0
2.717075768 0.000345266984716204
3.108458024 0.000598019965645366
3.626176704 0.000345266984716204
4.638408358 0
5.830579089 0.000488281254850639
7.774222389 0.000488281254850639
13.312624 0.000488281254850639
21.74411405 0.000488281254850639
35.77491095 0.000345266984716204
77.85465534 0.000488281254850639
145.4687624 0.000488281254850639
291.3761171 0.00133721337478927
};
\addlegendentry{NT=2000}
\addplot
table {%
2.324965044 0.000598019965645366
2.717075768 0.000772040463550879
3.108458024 0.000690533979722166
3.626176704 0.000690533979722166
4.638408358 0.000598019965645366
5.830579089 0.000690533979722166
7.774222389 0.000598019965645366
13.312624 0.000598019965645366
21.74411405 0.000690533979722166
35.77491095 0.000690533979722166
77.85465534 0.000598019965645366
145.4687624 0.000690533979722166
291.3761171 0.00158221224960986
};
\addlegendentry{NT=4000}
\end{axis}

\end{tikzpicture}
   \end{scaletikzpicturetowidth}
   
   \caption{Gradient computation: Angle between the lossy gradient vector and the reference gradient vector (in radians) vs \atol (left) and vs compression factor (right). If the lossy gradient vector was pointing in a significantly different direction as compared to the reference gradient, we could expect to see that on this plot. The angles are quite small. The number of timesteps do not affect the result by much. The results are also resilient to increasing \atol up to $10^{-2}$. Compression factors of over 100x do not seem to significantly distort the results either.}\label{fig:gradient_error_angle}
\end{figure}

Apart from showing that the error in the gradient remains almost constant with changing \atol, Figure~\ref{fig:gradient_error_angle} also shows that the number of timesteps do not appear to change the error by much (for a constant number of checkpoints). 

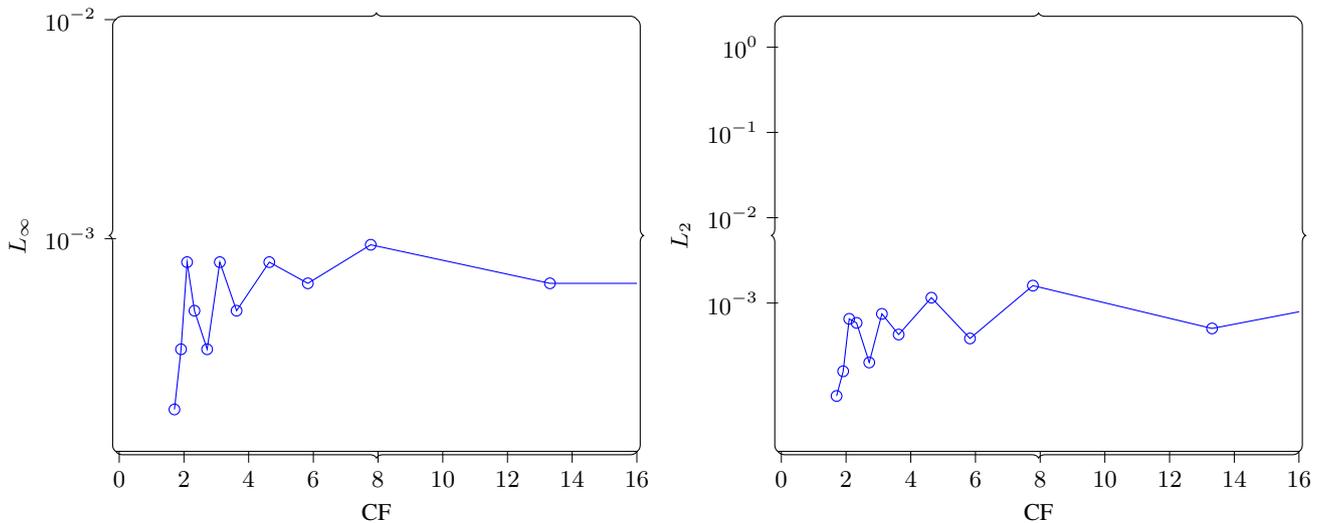
\begin{figure}[htbp]
\centering
   \begin{scaletikzpicturetowidth}{0.4\textwidth}
\begin{tikzpicture}

\begin{axis}[
cycle list name=cbw, 
log basis y={10},
tick align=outside,
tick pos=left,
x grid style={white!69.0196078431373!black},
xlabel={CF},
xmin=-0.1, xmax=16,
xtick style={color=black},
y grid style={white!69.0196078431373!black},
ylabel={$L_{\infty}$},
ymin=0.0106812983608182, ymax=1,
ymode=log,
ytick style={color=black},
ytick={0.1,1},
yticklabels={\(\displaystyle {10^{-3}}\),\(\displaystyle {10^{-2}}\),\(\displaystyle {10^{-1}}\),\(\displaystyle {10^{0}}\),\(\displaystyle {10^{1}}\),\(\displaystyle {10^{2}}\),\(\displaystyle {10^{3}}\),\(\displaystyle {10^{4}}\)}
]
\addplot
table {%
291.3761171 112.36914
145.4687624 15.33252
77.85465534 2.953125
35.77491095 0.18164062
21.74411405 0.0625
13.312624 0.0625
7.774222389 0.09375
5.830579089 0.0625
4.638408358 0.078125
3.626176704 0.046875
3.108458024 0.078125
2.717075768 0.03125
2.324965044 0.046875
2.097305367 0.078125
1.909927969 0.03125
1.706294796 0.016601562
};
\end{axis}

\end{tikzpicture}
\begin{tikzpicture}

\begin{axis}[
cycle list name=cbw, 
log basis y={10},
tick align=outside,
tick pos=left,
x grid style={white!69.0196078431373!black},
xlabel={CF},
xmin=-0.1, xmax=16,
xtick style={color=black},
y grid style={white!69.0196078431373!black},
ylabel={$L_2$},
ymin=0.0299702743229053, ymax=1,
ymode=log,
ytick style={color=black},
ytick={0.1,0.2, 0.4, 0.8},
yticklabels={\(\displaystyle {10^{-3}}\),\(\displaystyle {10^{-2}}\),\(\displaystyle {10^{-1}}\),\(\displaystyle {10^{0}}\),\(\displaystyle {10^{1}}\),\(\displaystyle {10^{2}}\),\(\displaystyle {10^{3}}\),\(\displaystyle {10^{4}}\)}
]
\addplot
table {%
145.4687624 51.08942
77.85465534 7.383107
35.77491095 0.5107779
21.74411405 0.12439996
13.312624 0.08136184
7.774222389 0.11520634
5.830579089 0.07501724
4.638408358 0.10446002
3.626176704 0.07743759
3.108458024 0.09159031
2.717075768 0.06169891
2.324965044 0.08511763
2.097305367 0.08795767
1.909927969 0.057456594
1.706294796 0.047001526
};
\end{axis}

\end{tikzpicture}
   \end{scaletikzpicturetowidth}
  
     \caption{Gradient error: $L_\infty$ (left) and $L_2$ (right) norms of the gradient error as a function of the achieved compression factor (CF). It can be seen that error is negligible in the range of CF up to 16. Compare this to subsampling in Figure~\ref{fig:subsampling}. Note that we achieved much higher CF values as part of the experiment but cut the axis in this figure to make it comparable to Figure~\ref{fig:subsampling}}\label{fig:gradient_error_cf}
\end{figure}

It can be seen from the plots that the errors induced in the checkpoint compression do not propagate significantly until the gradient computation step. In fact, the \atol compression setting does not affect the error in the gradient computation until a cutoff point. It is likely that the cross-correlation step in the gradient computation is acting as an error-correcting step since the adjoint computation continues at the same precision as before - the only errors introduced are in the values from the forward computation used in the cross-correlation step (the dotted arrows in Figure~\ref{fig:title}).

\begin{figure}[htbp]
\centering
   \begin{scaletikzpicturetowidth}{0.4\textwidth}
\begin{tikzpicture}

\begin{axis}[
cycle list name=cbw, 
legend cell align={left},
legend style={fill opacity=0.8, draw opacity=1, text opacity=1, at={(0.03,0.97)}, anchor=north west, draw=white!80!black},
log basis x={2},
log basis y={2},
tick align=outside,
tick pos=left,
title={},
x grid style={white!69.0196078431373!black},
xlabel={H},
xmin=0.00634572184653309, xmax=0.615572206672458,
xmode=log,
xtick style={color=black},
xtick={0.001953125,0.00390625,0.0078125,0.015625,0.03125,0.0625,0.125,0.25,0.5,1,2},
xticklabels={\(\displaystyle {2^{-9}}\),\(\displaystyle {2^{-8}}\),\(\displaystyle {2^{-7}}\),\(\displaystyle {2^{-6}}\),\(\displaystyle {2^{-5}}\),\(\displaystyle {2^{-4}}\),\(\displaystyle {2^{-3}}\),\(\displaystyle {2^{-2}}\),\(\displaystyle {2^{-1}}\),\(\displaystyle {2^{0}}\),\(\displaystyle {2^{1}}\)},
y grid style={white!69.0196078431373!black},
ylabel={error},
ymin=2.9059714014892, ymax=12182.9225687021,
ymode=log,
ytick style={color=black},
ytick={0.5,2,8,32,128,512,2048,8192,32768,131072},
yticklabels={\(\displaystyle {2^{-1}}\),\(\displaystyle {2^{1}}\),\(\displaystyle {2^{3}}\),\(\displaystyle {2^{5}}\),\(\displaystyle {2^{7}}\),\(\displaystyle {2^{9}}\),\(\displaystyle {2^{11}}\),\(\displaystyle {2^{13}}\),\(\displaystyle {2^{15}}\),\(\displaystyle {2^{17}}\)}
]
\addplot
table {%
0.0078125 4.24569327048187
0.015625 17.1497428294117
0.0312 67.6126225869921
0.0625 266.694099205671
0.125 1019.83844541597
0.25 3552.89732872558
0.5 8318.55703234958
};
\addlegendentry{Reference solution}
\addplot
table {%
0.0078125 4.55918358809822
0.015625 17.7767234646444
0.0312 68.864577519425
0.0625 269.202021746602
0.125 1024.85429049783
0.25 3562.92901888931
0.5 8338.62041267703
};
\addlegendentry{\atol$=10^{-1}$}
\end{axis}

\end{tikzpicture}
   \end{scaletikzpicturetowidth}
   
      \caption{Gradient linearization: Comparison of gradient linearization errors for \atol$=10^{-1}$ (left) and \atol$=10^{-2}$ (right) vs reference solution. The horizontal axis represents a small linear perturbation to the velocity model and the vertical axis represents the error observed at that perturbation. The two curves in each of the plots follow each other so closely that they are indistinguishable. This confirms that the lossy gradient satisfies the Taylor linearization properties just as well as the reference gradient. }\label{fig:gradient_linearization}
\end{figure}
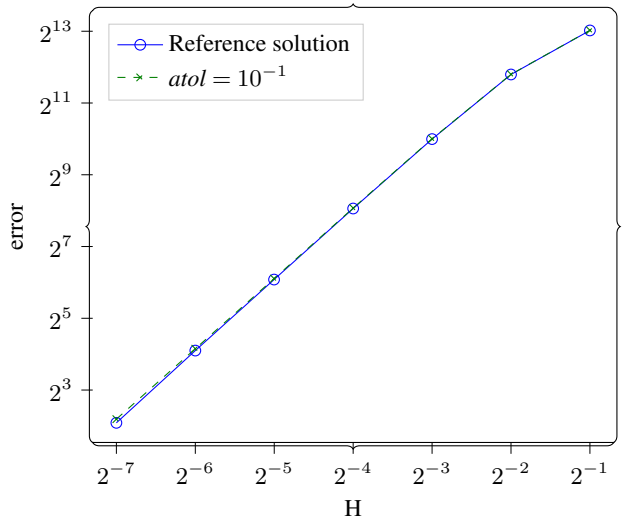

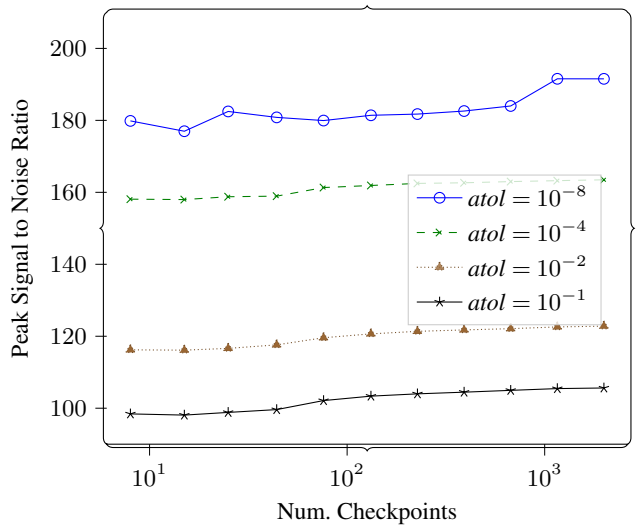
\begin{figure}[htbp]
   \centering
   
\begin{tikzpicture}

\begin{axis}[
cycle list name=cbw, 
legend style={fill opacity=0.8, draw opacity=1, text opacity=1, at={(0.95,0.45)}, anchor=east, draw=white!80!black},
log basis x={10},
tick align=outside,
tick pos=left,
x grid style={white!69.0196078431373!black},
xlabel={Num. Checkpoints},
xmin=6.07081979201724, xmax=2628.96948794073,
xmode=log,
xtick style={color=black},
xtick={0.1,1,10,100,1000,10000,100000},
xticklabels={\(\displaystyle {10^{-1}}\),\(\displaystyle {10^{0}}\),\(\displaystyle {10^{1}}\),\(\displaystyle {10^{2}}\),\(\displaystyle {10^{3}}\),\(\displaystyle {10^{4}}\),\(\displaystyle {10^{5}}\)},
y grid style={white!69.0196078431373!black},
ylabel={Peak Signal to Noise Ratio},
ymin=90, ymax=210,
ytick style={color=black}
]
\addplot
table {%
5 inf
8 179.85141564694
15 176.980189023265
25 182.478391389729
44 180.826889376249
76 179.945760969688
132 181.408330994921
227 181.745313491053
391 182.588740596815
673 183.998681269838
1159 191.551463905772
1995 191.540319366651
};
\addlegendentry{\atol$=10^{-8}$}
\addplot
table {%
5 inf
8 158.096400315477
15 157.984986691965
25 158.775611898309
44 158.938571224407
76 161.326574742352
132 161.900730558235
227 162.474662114044
391 162.652480417199
673 162.983337444851
1159 163.243054366037
1995 163.48367079302
};
\addlegendentry{\atol$=10^{-4}$}
\addplot
table {%
1 inf
5 inf
8 116.221278512643
15 116.133369472668
25 116.629547127811
44 117.601375338986
76 119.588234984725
132 120.661420150554
227 121.364261393015
391 121.757549169681
673 122.108890236568
1159 122.589319959873
1995 122.810246965514
};
\addlegendentry{\atol$=10^{-2}$}
\addplot 
table {%
5 inf
8 98.4170951714645
15 98.0868072350166
25 98.830254803285
44 99.6151025431769
76 102.124950962025
132 103.354248133841
227 103.991734391464
391 104.457929109031
673 104.962273620553
1159 105.45318739915
1995 105.621033253722
};
\addlegendentry{\atol$=10^{-1}$}
\end{axis}

\end{tikzpicture}
   \caption{Gradient error: In this plot we measure the effect of varying number of checkpoints on the error in the gradient. We report PSNR of lossy vs reference gradient as a function of number of checkpoints, for four different compression settings.}
   \label{fig:gradient_ncp}
\end{figure}

\subsection{Stacking}
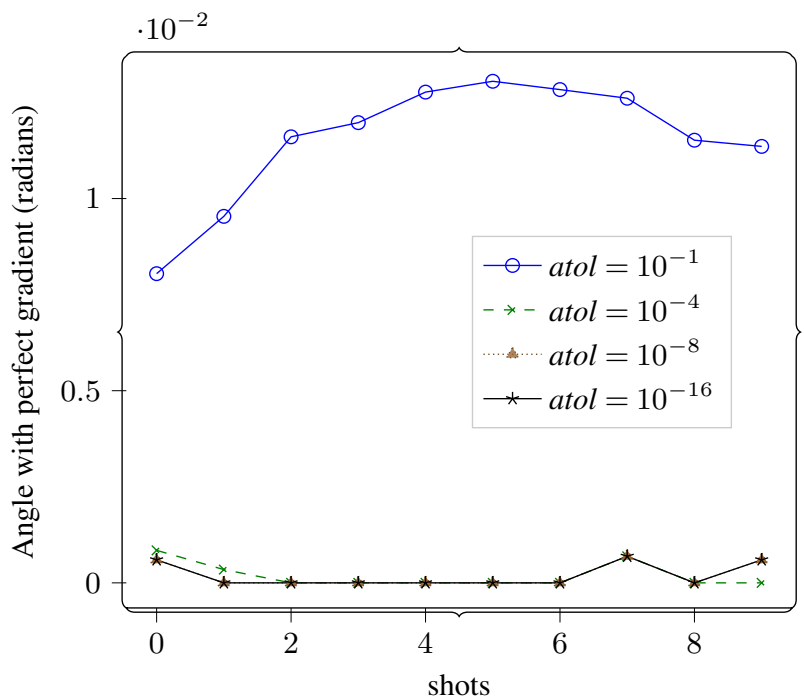
\begin{figure}[htbp]
    \centering
    \resizebox{0.6\textwidth}{!}{
\begin{tikzpicture}

\begin{axis}[
cycle list name=cbw,
legend cell align={left},
legend style={fill opacity=0.8, draw opacity=1, text opacity=1, at={(0.91,0.5)}, anchor=east, draw=white!80!black},
tick align=outside,
tick pos=left,
x grid style={white!69.0196078431373!black},
xlabel={shots},
xmin=-0.45, xmax=9.45,
xtick style={color=black},
y grid style={white!69.0196078431373!black},
ylabel={Angle with perfect gradient (radians)},
ymin=-0.00065259576516655, ymax=0.0137045110684975,
ytick style={color=black}
]
\addplot
table {%
0 0.00804555737040302
1 0.00953715771926821
2 0.0116063779609708
3 0.0119754118593933
4 0.0127702985229349
5 0.013051915303331
6 0.0128354781762597
7 0.0126106026708907
8 0.0115187415238455
9 0.0113571932154698
};
\addlegendentry{\atol $=10^{-1}$}
\addplot
table {%
0 0.000845727958587899
1 0.000345266984716204
2 0
3 0
4 0
5 0
6 0
7 0.000690533979722166
8 0
9 0
};
\addlegendentry{\atol $=10^{-4}$}
\addplot
table {%
0 0.000598019965645366
1 0
2 0
3 0
4 0
5 0
6 0
7 0.000690533979722166
8 0
9 0.000598019965645366
};
\addlegendentry{\atol $=10^{-8}$}
\addplot
table {%
0 0.000598019965645366
1 0
2 0
3 0
4 0
5 0
6 0
7 0.000690533979722166
8 0
9 0.000598019965645366
};
\addlegendentry{\atol $=10^{-16}$}
\end{axis}

\end{tikzpicture}}
    \caption{Shot stacking: The gradient is first computed for each individual shot and then added up for all the shots. In this experiment we measure the propagation of errors through this step. This plot shows that while errors do have the potential to accumulate through the step - as can be seen from the curve for $atol=10^{-1}$, for compression settings that are useful otherwise, the errors do not accumulate significantly. }
    \label{fig:stacking}
\end{figure}

After gradient computation on a single shot, the next step in FWI is the accumulation of the gradients for individual shots by adding them into a single gradient. We call this stacking. In this experiment we studied the accumulation of errors through this stacking process. Figure~\ref{fig:stacking} shows the error in the gradient computation (compared to a similarly processed reference problem) as a function of the number of shots. 

This plot shows us that the errors across the different shots are not adding up and the cumulative error is not growing with the number of shots - except for the compression setting of $atol=10^{-1}$, which is chosen as an example of unreasonably high compression. 

\subsection{Convergence}

\begin{figure}[htbp]
\centering
   \includegraphics[width=\textwidth]{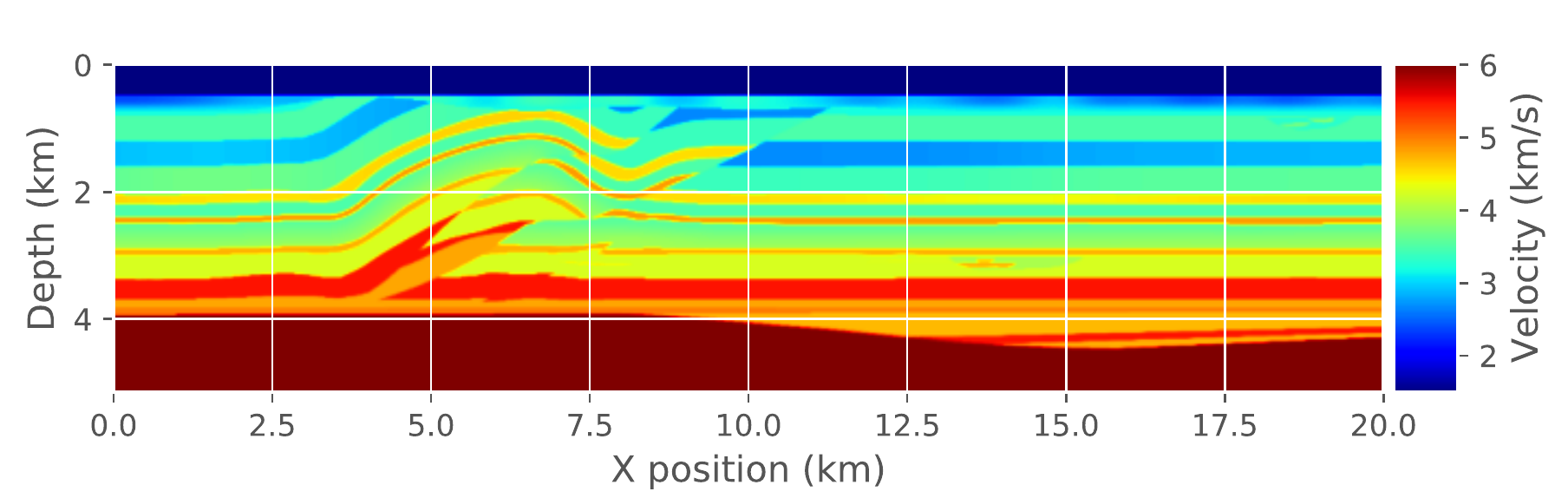} 
      \caption{True solution for FWI}\label{fig:fwi_reference_true}
\end{figure}

\begin{figure}[htbp]
\centering   
   \includegraphics[width=\textwidth]{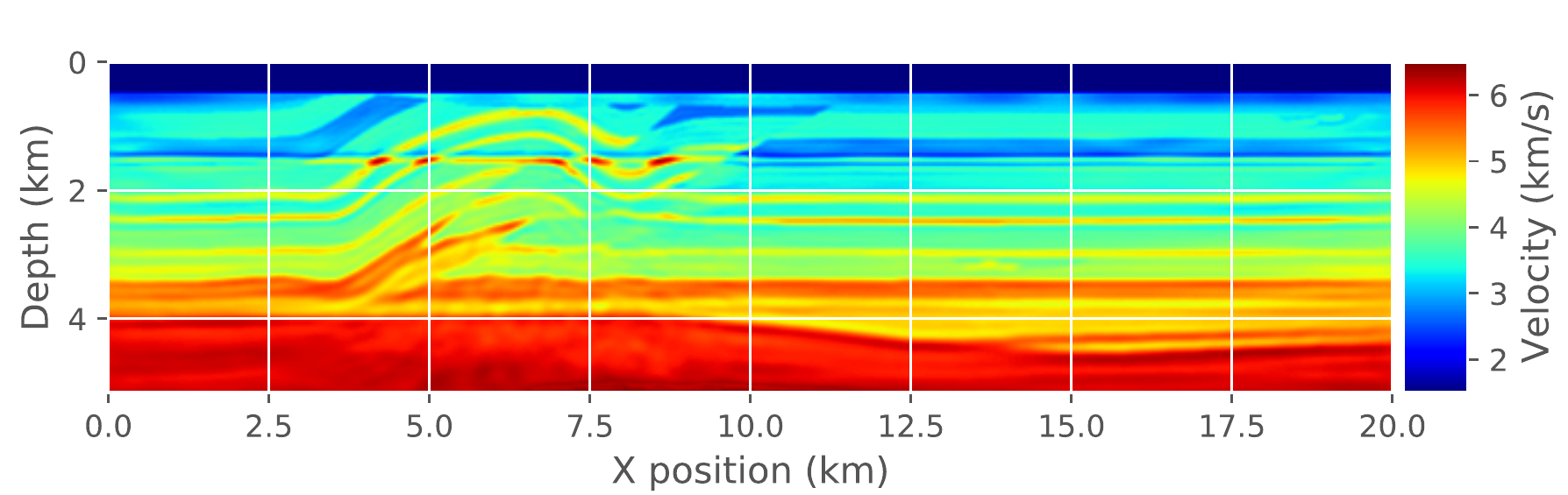}

   \caption{Reference solution for the complete FWI problem. This is the solution after running \emph{reference} FWI for 30 iterations}\label{fig:fwi_reference}
\end{figure}

\begin{figure}[htbp]
   \centering
   \includegraphics[width=\textwidth]{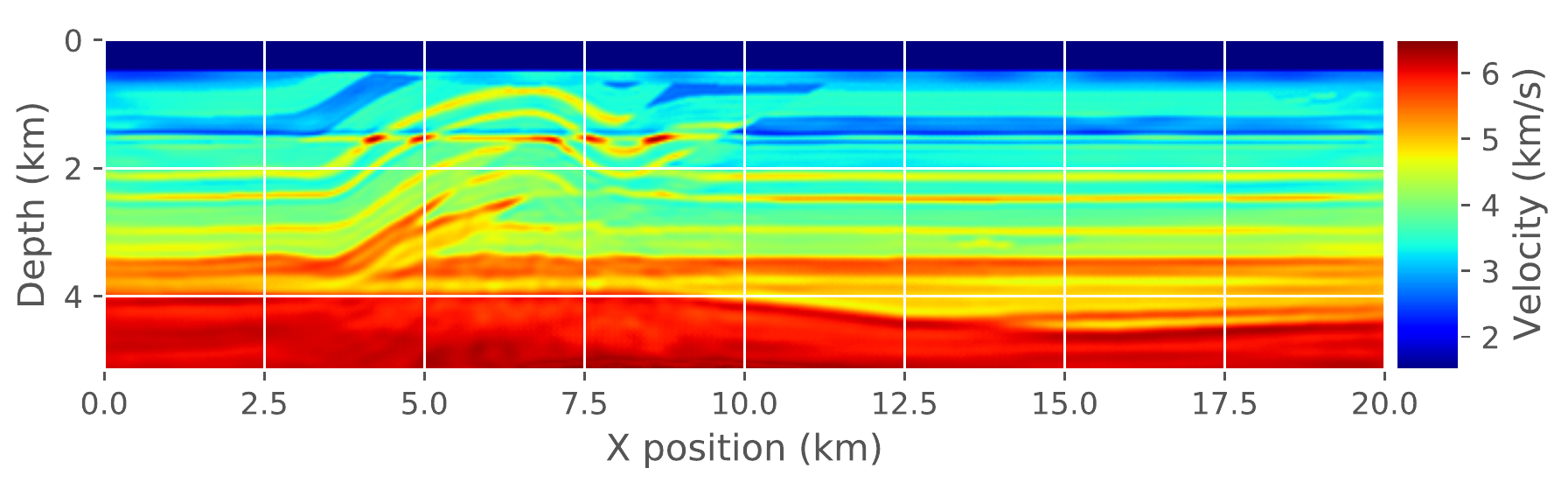} 
   \caption{Final image after running FWI \atol$=10^{-1}$. It is visually indistinguishable from the reference solution in Figure~\ref{fig:fwi_reference}.}
   \label{fig:fwi_final}
\end{figure}

Finally, we measure the effect of an approximate gradient on the convergence of the FWI problem. For reference, Figure~\ref{fig:fwi_reference} shows the known true velocity model for this problem. Figure~\ref{fig:fwi_reference} shows the final velocity model after running a \emph{reference} FWI for 30 iterations. Figure~\ref{fig:fwi_final} shows the final velocity model after running FWI with compression enabled at different \atol settings - also for 30 iterations. 

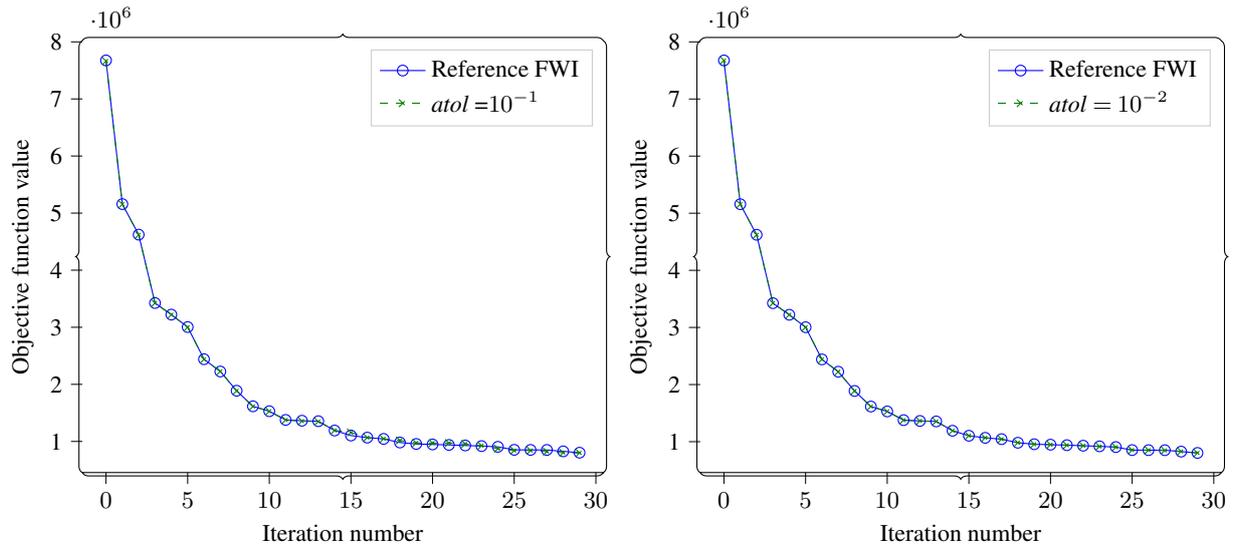
\begin{figure}[htbp]
   \centering
\begin{tikzpicture}

\definecolor{color0}{rgb}{0.12156862745098,0.466666666666667,0.705882352941177}
\definecolor{color1}{rgb}{1,0.498039215686275,0.0549019607843137}

\begin{axis}[
cycle list name=cbw, 
legend cell align={left},
legend style={fill opacity=0.8, draw opacity=1, text opacity=1, draw=white!80!black},
tick align=outside,
tick pos=left,
title={},
x grid style={white!69.0196078431373!black},
xlabel={Iteration number},
xmin=-1.45, xmax=30.45,
xtick style={color=black},
y grid style={white!69.0196078431373!black},
ylabel={Objective function value},
ymin=454755.097613722, ymax=8018884.98165392,
ytick style={color=black},
ytick={0,1000000,2000000,3000000,4000000,5000000,6000000,7000000,8000000,9000000},
yticklabels={0,1,2,3,4,5,6,7,8,9}
]
\addplot 
table {%
0 7675060.89601573
1 5157783.31949366
2 4622413.60209328
3 3425058.20491496
4 3221846.9954475
5 3003643.54973162
6 2440358.03002281
7 2224985.37653164
8 1888026.1594263
9 1615325.20288123
10 1529443.51831116
11 1377034.46622922
12 1363035.53429328
13 1354565.8016982
14 1191268.033823
15 1103720.32664367
16 1067170.05477961
17 1044122.23786061
18 981185.940108711
19 955585.857799024
20 947462.151554451
21 938688.812264019
22 929563.152725196
23 918597.42729378
24 905534.665492946
25 852687.731216872
26 851714.800487069
27 850742.011234269
28 826216.599514826
29 802826.49574388
};
\addlegendentry{Reference FWI}
\addplot 
table {%
0 7663376.81384267
1 5155197.16449903
2 4618634.66302407
3 3422883.67395488
4 3219795.36732561
5 3001105.97773512
6 2438256.97332003
7 2228358.59272185
8 1886571.02544311
9 1613793.17837746
10 1528489.65323916
11 1375925.08449575
12 1355523.60290228
13 1347938.55707426
14 1193945.00379757
15 1157005.20926087
16 1067350.97690593
17 1046636.1692485
18 1018037.59449166
19 973639.581460998
20 971587.469925011
21 971347.760536634
22 955657.132360857
23 926235.184221873
24 873445.967465233
25 841761.671579384
26 841168.460844874
27 822209.464932268
28 802539.730212883
29 798579.183251913
};
\addlegendentry{\atol=$10^{-1}$}
\end{axis}

\end{tikzpicture}
\begin{tikzpicture}

\begin{axis}[
cycle list name=cbw, 
legend cell align={left},
legend style={fill opacity=0.8, draw opacity=1, text opacity=1, draw=white!80!black},
tick align=outside,
tick pos=left,
title={},
x grid style={white!69.0196078431373!black},
xlabel={Iteration number},
xmin=-1.45, xmax=30.45,
xtick style={color=black},
y grid style={white!69.0196078431373!black},
ylabel={Objective function value},
ymin=459214.775730287, ymax=8018672.61602932,
ytick style={color=black},
ytick={0,1000000,2000000,3000000,4000000,5000000,6000000,7000000,8000000,9000000},
yticklabels={0,1,2,3,4,5,6,7,8,9}
]
\addplot
table {%
0 7675060.89601573
1 5157783.31949366
2 4622413.60209328
3 3425058.20491496
4 3221846.9954475
5 3003643.54973162
6 2440358.03002281
7 2224985.37653164
8 1888026.1594263
9 1615325.20288123
10 1529443.51831116
11 1377034.46622922
12 1363035.53429328
13 1354565.8016982
14 1191268.033823
15 1103720.32664367
16 1067170.05477961
17 1044122.23786061
18 981185.940108711
19 955585.857799024
20 947462.151554451
21 938688.812264019
22 929563.152725196
23 918597.42729378
24 905534.665492946
25 852687.731216872
26 851714.800487069
27 850742.011234269
28 826216.599514826
29 802826.49574388
};
\addlegendentry{Reference FWI}
\addplot
table {%
0 7674730.09957926
1 5157693.09284474
2 4622286.07715702
3 3424959.71963781
4 3221760.04695115
5 3003564.4716356
6 2440302.30144199
7 2224894.97065917
8 1887977.89597634
9 1615216.45180235
10 1529382.8807264
11 1376960.1361706
12 1363091.09552189
13 1354735.41556023
14 1191210.21556989
15 1103808.1742542
16 1067312.18303686
17 1044099.05720185
18 981614.691888457
19 955672.937764359
20 947237.452570592
21 938107.880393557
22 928310.267742916
23 917142.078890549
24 903659.77872607
25 852248.363134789
26 851043.610058413
27 847554.222632047
28 829770.385867127
29 803213.712244284
};
\addlegendentry{\atol$=10^{-2}$}
\end{axis}

\end{tikzpicture}
      \caption{Convergence: As the last experiment, we run complete FWI to convergence (up to max 30 iterations). Here we show the convergence profiles for \atol$=10^{-1}$ (left) and \atol$=10^{-2}$ (right) vs the reference problem. The reference curve is so closely followed by the lossy curve that the reference curve is hidden behind.}
   \label{fig:fwi_convergence}
\end{figure}

Figure~\ref{fig:fwi_convergence} shows the convergence trajectory - the objective function value as a function of the iteration number. We show this convergence trajectory for 4 different compression settings. It can be seen that the compressed version does indeed follow a very similar trajectory as the original problem.

\begin{figure}[htbp]
   \centering
   \includegraphics[width=\textwidth]{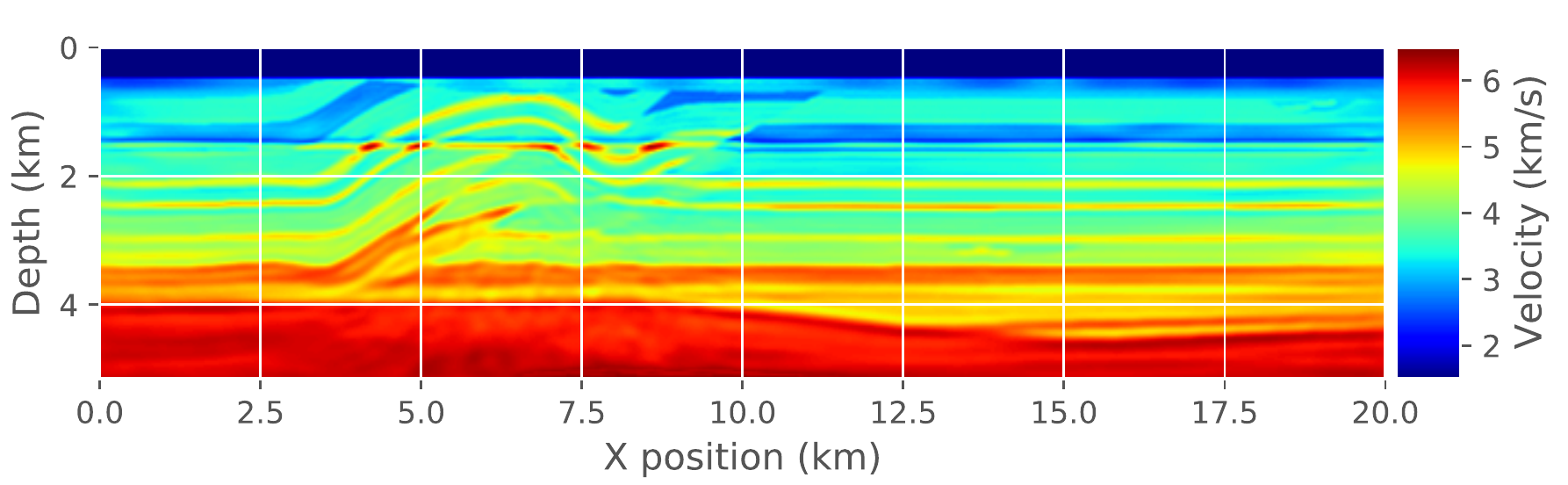}
   \caption{Final image after running FWI \atol$=10^{-2}$. It is visually indistinguishable from the reference solution in Figure~\ref{fig:fwi_reference}.}
   \label{fig:fwi_final_2}
\end{figure}

\subsection{Subsampling}
\label{sec:subsampling}
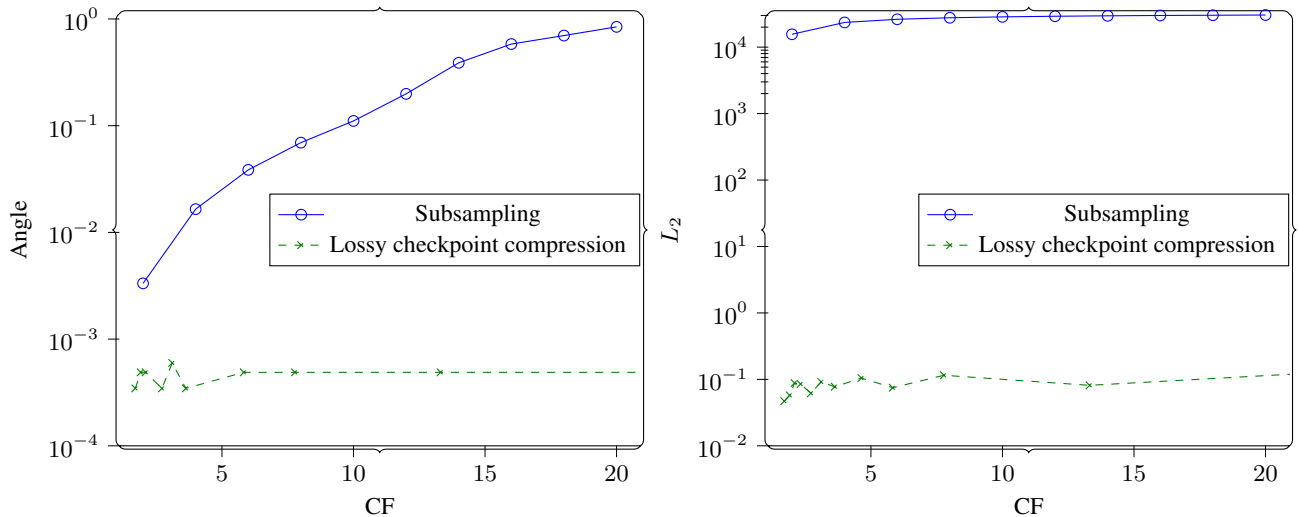
\begin{figure}[htbp]
\centering

   \begin{scaletikzpicturetowidth}{0.4\textwidth}
\begin{tikzpicture}

\begin{axis}[
cycle list name=cbw,
log basis y={10},
tick align=outside,
tick pos=left,
x grid style={white!69.0196078431373!black},
xlabel={CF},
xmin=1.1, xmax=20.9,
xtick style={color=black},
y grid style={white!69.0196078431373!black},
ylabel={Angle},
ymin=0.0001, ymax=1.11208609927805,
ymode=log,
ytick style={color=black},
ytick={0.0001,0.001,0.01,0.1,1,10,100},
yticklabels={\(\displaystyle {10^{-4}}\),\(\displaystyle {10^{-3}}\),\(\displaystyle {10^{-2}}\),\(\displaystyle {10^{-1}}\),\(\displaystyle {10^{0}}\),\(\displaystyle {10^{1}}\),\(\displaystyle {10^{2}}\)},
legend style={at={(1,0.5)},anchor=east}
]
\addplot
table {%
2 0.0033296357414455
4 0.0165153581292294
6 0.0385395013127642
8 0.069370474736466
10 0.110764629668043
12 0.198779045407259
14 0.388448916229751
16 0.581921692957683
18 0.699049959277501
20 0.843257546828501
};
\addlegendentry{Subsampling}
\addplot
table {%
1.706294796 0.000345266984716204
1.909927969 0.000488281254850639
2.097305367 0.000488281254850639
2.324965044 0
2.717075768 0.000345266984716204
3.108458024 0.000598019965645366
3.626176704 0.000345266984716204
4.638408358 0
5.830579089 0.000488281254850639
7.774222389 0.000488281254850639
13.312624 0.000488281254850639
21.74411405 0.000488281254850639
};
\addlegendentry{Lossy checkpoint compression}
\end{axis}

\end{tikzpicture}
\begin{tikzpicture}

\begin{axis}[
cycle list name=cbw,
log basis y={10},
minor ytick={2000,3000,4000,5000,6000,7000,8000,9000,20000,30000,40000,50000,60000,70000,80000,90000,200000,300000,400000,500000,600000,700000,800000,900000,2000000,3000000,4000000,5000000,6000000,7000000,8000000,9000000},
tick align=outside,
tick pos=left,
x grid style={white!69.0196078431373!black},
xlabel={CF},
xmin=1.1, xmax=20.9,
xtick style={color=black},
y grid style={white!69.0196078431373!black},
ylabel={$L_2$},
ymin=0.01, ymax=31464.1030719384,
ymode=log,
ytick style={color=black},
ytick={0.01, 0.1, 1, 10, 100, 1000,10000,100000,1000000},
yticklabels={\(\displaystyle {10^{-2}}\), \(\displaystyle {10^{-1}}\), \(\displaystyle {10^{0}}\), \(\displaystyle {10^{1}}\), \(\displaystyle {10^{2}}\), \(\displaystyle {10^{3}}\),\(\displaystyle {10^{4}}\),\(\displaystyle {10^{5}}\),\(\displaystyle {10^{6}}\)},
legend style={at={(1,0.5)},anchor=east}
]
\addplot
table {%
2 15618.897
4 23530.502
6 26252.217
8 27668.088
10 28552.783
12 29162.52
14 29602.432
16 29958.29
18 30202.441
20 30432.057
};
\addlegendentry{Subsampling}
\addplot
table {%
21.74411405 0.12439996
13.312624 0.08136184
7.774222389 0.11520634
5.830579089 0.07501724
4.638408358 0.10446002
3.626176704 0.07743759
3.108458024 0.09159031
2.717075768 0.06169891
2.324965044 0.08511763
2.097305367 0.08795767
1.909927969 0.057456594
1.706294796 0.047001526
};
\addlegendentry{Lossy checkpoint compression}
\end{axis}

\end{tikzpicture}
   \end{scaletikzpicturetowidth}
   
   \caption{Subsampling: We set up an experiment with subsampling as a baseline for comparison. Subsampling is when the gradient computation is carried at a lower timestepping than the simulation itself. This requires less data to be carried over from the forward to the reverse computation at the cost of solution accuracy so is comparable to lossy checkpoint compression. This plot shows the angle between the lossy gradient and the reference gradient versus the compression factor $CF$ (left) and $L_2$ norm of gradient error versus the compression factor $CF$ (right) for this experiment. Compare this to the errors in Figure~\ref{fig:gradient_error_cf} that apply for lossy checkpoint compression.}\label{fig:subsampling}
\end{figure}

To compare our proposed method with subsampling - which is sometimes used in industry, we run an experiment where we use subsampling in time to reduce the memory footprint. Figure~\ref{fig:subsampling} shows some error metrics as a function of the compression factor $f$. 

Comparing Figure~\ref{fig:subsampling} with Figure~\ref{fig:gradient_error_cf}, it can be seen that the proposed method produces significantly smaller errors for similar compression factors. 

\section{Discussion}

The results indicate that significant lossy compression can be applied to the checkpoints before the solution is adversary affected. This is an interesting result because, while it is common to see approximate methods leading to approximate solutions, this is not what we see in our results - the solution error doesn't change much for large compression error. This being an empirical study, we can only speculate on the reasons for this. We know that in the proposed method, the adjoint computation is not affected at all - the only effect is in the wavefield carried over from the forward computation to the gradient computation step. Since the gradient computation is a cross-correlation, we only expect correlated signals to grow in magnitude in the gradient calculation and when gradients are stacked. The optimization steps are likely to be error-correcting as well since even with an approximate gradient ($atol > 4$), the convergence trajectory and the final results do not appear to change much - indicating that the errors in the gradient might be canceling out over successive iterations. There is even the possibility that these \emph{errors} in the gradient introduce a kind of regularization \citep{zchecker}. This is likely since we know from \citet{zfperror}, that ZFP's errors are likely to \emph{smoothen} the field being compressed. We also know from \citet{zchecker} that ZFP's errors are likely to be (close-to) normally distributed with 0 mean, which reinforces the idea that this is likely to act as a regularizer. We only tried this with the ZFP compression algorithm in this work. ZFP being a generic floating-point compression algorithm, is likely to be more broadly applicable than application-specific compressors. This makes it more broadly useful for Devito, which was the DSL that provided the context for this work. Some clear choices for the next compressors to try would be SZ \citep{di2016fast} - which is also a generic floating-point compression library, and the application-specific compressors from \citet{weiser2012state, boehm2016wavefield, marin2016large}. A different compression algorithm would change: 
\begin{itemize}
    \item the error distribution, 
    \item the compression/decompression times, and
    \item the achieved compression factors.
\end{itemize}
Based on the experiments from \citet{zchecker}, we would expect the errors in SZ to be (nearly) uniformly distributed with a 0 mean. It would be interesting to see the effect this new distribution has on the method we describe here. If a new compressor can achieve higher compression factors than ZFP (for illustration), in less compression/decompression time than ZFP, then it will clearly speed up the application relative to ZFP.  In reality, the relationship is likely to be more complex, and the performance model from \citet{kukreja2019combining} helps compare the performance of various compressors on this problem without running the full problem. The number of checkpoints has some effect on the error - more checkpoints incur less error for the same compression setting - as would be expected. Since we showed the benefits of compression for inversion, the expected speedup does not depend on the medium that varies between iterations from extremely smooth to very heterogeneous. While we focused on acoustic waves in this work for simplicity, different physics should not impact the compression factor due to the strong similarity between the solutions of different wave equations. However, different physics might require more fields in the solution - increasing the memory requirements, while also increasing the computational requirements. Whether this increase favours compression or recomputation more depends on the operational intensity of the specific wave equation kernel.  The choice of misfit function is also not expected to impact our results since the wavefield does not depend on the misfit function. A more thorough study will, however, be necessary to generalize our results to other problem domains such as computational fluid dynamics that involve drastically different solutions.

Our method accepts an acceptable error tolerance as input for every gradient evaluation. We expect this to be provided as part of an adaptive optimization scheme that requires approximate gradients in the first few iterations of the optimization, and progressively more accurate gradients as the solution approaches the optimum. Such a scheme was previously described in \citet{blanchet2019convergence}. Implementing such an adaptive optimizer and using it in practice is ongoing work.

\conclusions[Conclusions and Future Work]
In the preceding sections, we have shown that using lossy compression, high compression factors can be achieved without significantly impacting the convergence or final solution of the inversion solver. This is a very promising result for the use of lossy compression in FWI. The use of compression in large computations like this is especially important in the exascale era, where the gap between computing and memory speed is increasingly large. Compression can reduce the strain on the memory bandwidth by trading it off for extra computation - this is especially useful since modern CPUs are hard to saturate with low OI computations. 

In future work, we would like to study the interaction between compression errors and the velocity model for which FWI is being solved, as well as the source frequency. We would also like to compare multiple lossy compression algorithms e.g., \emph{SZ}.




\codedataavailability{The data used was from the Overthrust model, provided by SEG/EAGE \citep{fred_aminzadeh_1997_4252588}. The code for the scripts used here \citep{navjot_kukreja_2020_4247199}, the Devito DSL \citep{fabio_luporini_2020_3973710}, and pyzfp \citep{navjot_kukreja_2020_4252530} is all available online through Zenodo.} 



\appendix
\section{Additional Results}
\subsection{Direct compression}
\begin{figure}[H]
\centering

   \begin{scaletikzpicturetowidth}{0.1\textwidth}
\begin{tikzpicture}

\begin{axis}[
cycle list name=cbw, 
log basis x={10},
log basis y={10},
tick align=outside,
tick pos=left,
x grid style={white!69.0196078431373!black},
xlabel={atol},
xmin=1.58489319246111e-17, xmax=6.30957344480193,
xmode=log,
xtick style={color=black},
xtick={1e-19,1e-17,1e-15,1e-13,1e-11,1e-09,1e-07,1e-05,0.001,0.1,10,1000},
xticklabels={\(\displaystyle {10^{-19}}\),\(\displaystyle {10^{-17}}\),\(\displaystyle {10^{-15}}\),\(\displaystyle {10^{-13}}\),\(\displaystyle {10^{-11}}\),\(\displaystyle {10^{-9}}\),\(\displaystyle {10^{-7}}\),\(\displaystyle {10^{-5}}\),\(\displaystyle {10^{-3}}\),\(\displaystyle {10^{-1}}\),\(\displaystyle {10^{1}}\),\(\displaystyle {10^{3}}\)},
y grid style={white!69.0196078431373!black},
ylabel={$L_{\infty}$},
ymin=2.80214785474735e-18, ymax=0.650667611163033,
ymode=log,
ytick style={color=black},
ytick={1e-20,1e-18,1e-16,1e-14,1e-12,1e-10,1e-08,1e-06,0.0001,0.01,1,100},
yticklabels={\(\displaystyle {10^{-20}}\),\(\displaystyle {10^{-18}}\),\(\displaystyle {10^{-16}}\),\(\displaystyle {10^{-14}}\),\(\displaystyle {10^{-12}}\),\(\displaystyle {10^{-10}}\),\(\displaystyle {10^{-8}}\),\(\displaystyle {10^{-6}}\),\(\displaystyle {10^{-4}}\),\(\displaystyle {10^{-2}}\),\(\displaystyle {10^{0}}\),\(\displaystyle {10^{2}}\)},
cycle list name=cbw,
scale only axis
]
\addplot
table {%
1 0.105682127177715
0.1 0.0116252817679197
0.01 0.00141748040914536
0.001 0.000204496085643768
0.0001 1.55267771333456e-05
1e-05 2.06008553504944e-06
1e-06 2.68453732132912e-07
1e-07 1.76441972143948e-08
1e-08 2.3246684577316e-09
1e-09 2.97859514830634e-10
1e-10 1.96536120711244e-11
1e-11 2.97895041967422e-12
1e-12 3.61932706027801e-13
1e-13 2.59237076249974e-14
1e-14 2.21697660229836e-15
1e-15 2.65846372693446e-16
1e-16 1.72523670696756e-17
};
\end{axis}

\end{tikzpicture}
\begin{tikzpicture}

\begin{axis}[
cycle list name=cbw, 
log basis x={10},
log basis y={10},
tick align=outside,
tick pos=left,
x grid style={white!69.0196078431373!black},
xlabel={atol},
xmin=1.58489319246111e-17, xmax=6.30957344480193,
xmode=log,
xtick style={color=black},
xtick={1e-19,1e-17,1e-15,1e-13,1e-11,1e-09,1e-07,1e-05,0.001,0.1,10,1000},
xticklabels={\(\displaystyle {10^{-19}}\),\(\displaystyle {10^{-17}}\),\(\displaystyle {10^{-15}}\),\(\displaystyle {10^{-13}}\),\(\displaystyle {10^{-11}}\),\(\displaystyle {10^{-9}}\),\(\displaystyle {10^{-7}}\),\(\displaystyle {10^{-5}}\),\(\displaystyle {10^{-3}}\),\(\displaystyle {10^{-1}}\),\(\displaystyle {10^{1}}\),\(\displaystyle {10^{3}}\)},
y grid style={white!69.0196078431373!black},
ylabel={$L_1$},
ymin=2.33889424953295e-12, ymax=1954302.9374868,
ymode=log,
ytick style={color=black},
ytick={1e-15,1e-12,1e-09,1e-06,0.001,1,1000,1000000,1000000000,1000000000000},
yticklabels={\(\displaystyle {10^{-15}}\),\(\displaystyle {10^{-12}}\),\(\displaystyle {10^{-9}}\),\(\displaystyle {10^{-6}}\),\(\displaystyle {10^{-3}}\),\(\displaystyle {10^{0}}\),\(\displaystyle {10^{3}}\),\(\displaystyle {10^{6}}\),\(\displaystyle {10^{9}}\),\(\displaystyle {10^{12}}\)},
cycle list name=cbw,
scale only axis
]
\addplot
table {%
1 299472.102210334
0.1 67961.0935924164
0.01 14309.2959615832
0.001 2615.39478019412
0.0001 237.231240088658
1e-05 36.6077700149902
1e-06 5.86281899339779
1e-07 0.410100953736283
1e-08 0.052523445634208
1e-09 0.0066501171156702
1e-10 0.000388119945654918
1e-11 3.68973085912541e-05
1e-12 2.67346639865648e-06
1e-13 6.99535158842893e-08
1e-14 4.77217340098683e-09
1e-15 3.59536482304535e-10
1e-16 1.52632177374668e-11
};
\end{axis}

\end{tikzpicture}
   \end{scaletikzpicturetowidth}

   \caption{Direct compression: On the top, $L_{\infty}$ norm of error versus \atol. This plot verifies that ZFP respects the tolerance we set. On the bottom, $L_1$ norm of error versus \atol. From the difference in magnitude between the $L_{\infty}$ plot and this one, we can see how the error is spread across the domain.}\label{fig:direct_error_Linf}
\end{figure}
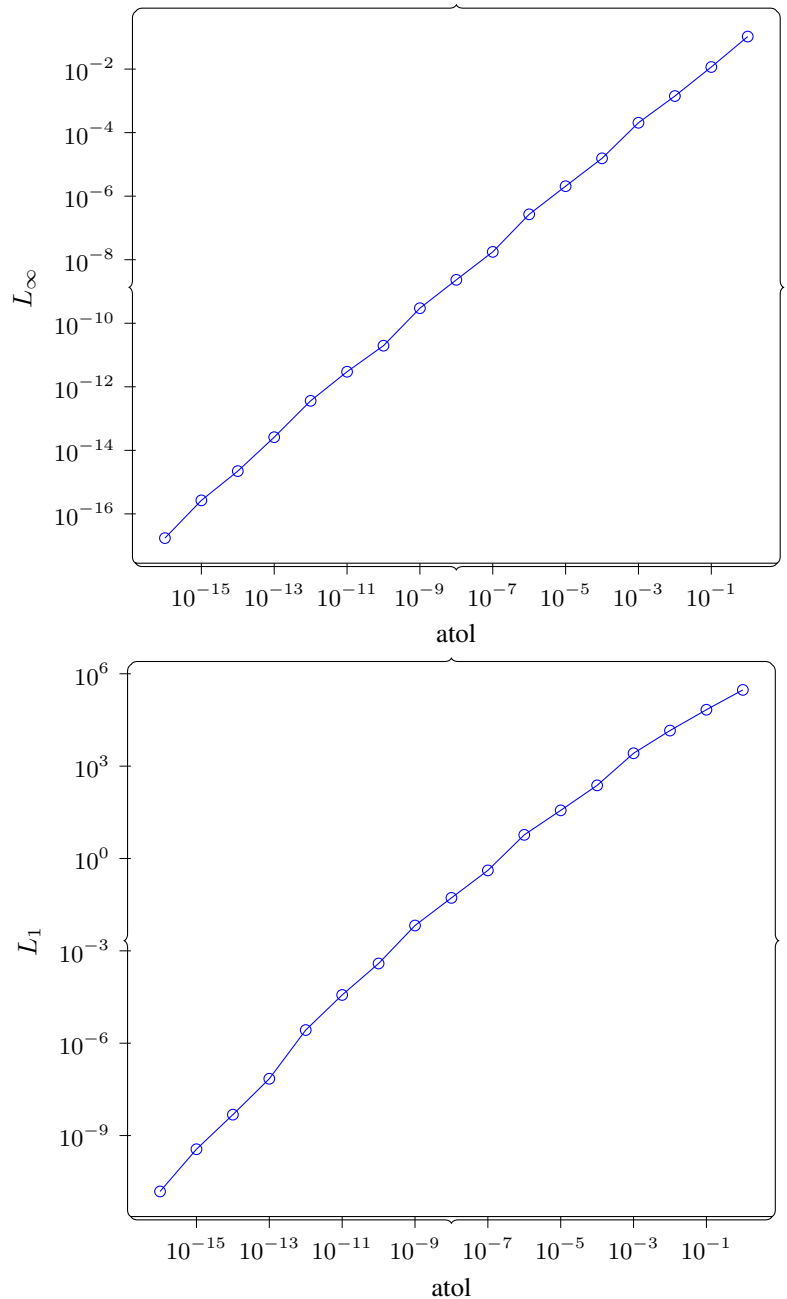

\subsection{Gradient Computation}
\begin{figure}[H]
\centering
   \begin{scaletikzpicturetowidth}{0.4\textwidth}
\begin{tikzpicture}

\begin{axis}[
cycle list name=cbw, 
log basis x={10},
log basis y={10},
tick align=outside,
tick pos=left,
x grid style={white!69.0196078431373!black},
xlabel={atol},
xmin=1.77827941003892e-17, xmax=0.562341325190349,
xmode=log,
xtick style={color=black},
xtick={1e-19,1e-17,1e-15,1e-13,1e-11,1e-09,1e-07,1e-05,0.001,0.1,10,1000},
xticklabels={\(\displaystyle {10^{-19}}\),\(\displaystyle {10^{-17}}\),\(\displaystyle {10^{-15}}\),\(\displaystyle {10^{-13}}\),\(\displaystyle {10^{-11}}\),\(\displaystyle {10^{-9}}\),\(\displaystyle {10^{-7}}\),\(\displaystyle {10^{-5}}\),\(\displaystyle {10^{-3}}\),\(\displaystyle {10^{-1}}\),\(\displaystyle {10^{1}}\),\(\displaystyle {10^{3}}\)},
y grid style={white!69.0196078431373!black},
ylabel={$L_{\infty}$},
ymin=0.0106812983608182, ymax=174.651356190913,
ymode=log,
ytick style={color=black},
ytick={0.001,0.01,0.1,1,10,100,1000,10000},
yticklabels={\(\displaystyle {10^{-3}}\),\(\displaystyle {10^{-2}}\),\(\displaystyle {10^{-1}}\),\(\displaystyle {10^{0}}\),\(\displaystyle {10^{1}}\),\(\displaystyle {10^{2}}\),\(\displaystyle {10^{3}}\),\(\displaystyle {10^{4}}\)}
]
\addplot
table {%
0.1 112.36914
0.01 15.33252
0.001 2.953125
0.0001 0.18164062
1e-05 0.0625
1e-06 0.0625
1e-07 0.09375
1e-08 0.0625
1e-09 0.078125
1e-10 0.046875
1e-11 0.078125
1e-12 0.03125
1e-13 0.046875
1e-14 0.078125
1e-15 0.03125
1e-16 0.016601562
};
\end{axis}

\end{tikzpicture}
\begin{tikzpicture}

\begin{axis}[
cycle list name=cbw, 
log basis x={10},
log basis y={10},
tick align=outside,
tick pos=left,
x grid style={white!69.0196078431373!black},
xlabel={atol},
xmin=1.77827941003892e-17, xmax=0.562341325190349,
xmode=log,
xtick style={color=black},
xtick={1e-19,1e-17,1e-15,1e-13,1e-11,1e-09,1e-07,1e-05,0.001,0.1,10,1000},
xticklabels={\(\displaystyle {10^{-19}}\),\(\displaystyle {10^{-17}}\),\(\displaystyle {10^{-15}}\),\(\displaystyle {10^{-13}}\),\(\displaystyle {10^{-11}}\),\(\displaystyle {10^{-9}}\),\(\displaystyle {10^{-7}}\),\(\displaystyle {10^{-5}}\),\(\displaystyle {10^{-3}}\),\(\displaystyle {10^{-1}}\),\(\displaystyle {10^{1}}\),\(\displaystyle {10^{3}}\)},
y grid style={white!69.0196078431373!black},
ylabel={L2},
ymin=0.0299702743229053, ymax=596.977556953662,
ymode=log,
ytick style={color=black},
ytick={0.001,0.01,0.1,1,10,100,1000,10000},
yticklabels={\(\displaystyle {10^{-3}}\),\(\displaystyle {10^{-2}}\),\(\displaystyle {10^{-1}}\),\(\displaystyle {10^{0}}\),\(\displaystyle {10^{1}}\),\(\displaystyle {10^{2}}\),\(\displaystyle {10^{3}}\),\(\displaystyle {10^{4}}\)}
]
\addplot
table {%
0.1 380.65958
0.01 51.08942
0.001 7.383107
0.0001 0.5107779
1e-05 0.12439996
1e-06 0.08136184
1e-07 0.11520634
1e-08 0.07501724
1e-09 0.10446002
1e-10 0.07743759
1e-11 0.09159031
1e-12 0.06169891
1e-13 0.08511763
1e-14 0.08795767
1e-15 0.057456594
1e-16 0.047001526
};
\end{axis}

\end{tikzpicture}
   \end{scaletikzpicturetowidth}
      \caption{Gradient computation: $L_\infty$ norm of gradient error versus \atol (left) and $L_2$ norm of gradient error versus \atol. It can be seen that the error stays almost constant and very low up to a threshold value of $10^{-4}$}\label{gradient_error}
\end{figure}
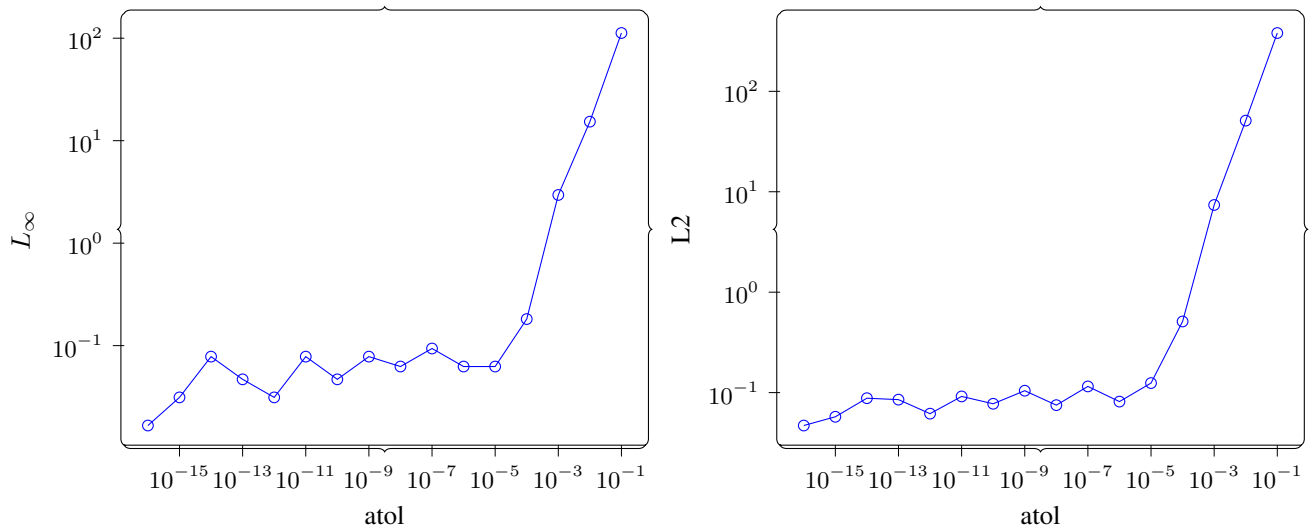

\noappendix       




\appendixfigures  

\appendixtables   


\authorcontribution{Most of the code and experimentation were done by Navjot. The experiments were planned between Jan and Navjot. Mathias helped set up meaningful experiments. John contributed in finetuning the experiments and the presentation of results. Paul and Gerard gave the overall direction of the work. Everybody contributed to the writing.} 

\competinginterests{The authors have no competing interests to declare} 


\begin{acknowledgements}
This work was funded in part by support from the U.S. Department of Energy, Office of Science, under contract DE-AC02-06CH11357. This research was carried out with the support of Georgia Research Alliance and partners of the ML4Seismic Center.
\end{acknowledgements}



\bibliographystyle{copernicus}
\bibliography{references.bib}

\end{document}